\documentclass[twocolumn,aps,reprint]{revtex4-2}
\pdfoutput=1
\usepackage[T1]{fontenc}
\usepackage{epsfig,amsmath}
\usepackage{amssymb}
\usepackage{subfigure}
\usepackage{graphicx}
\usepackage[colorlinks,linkcolor=blue]{hyperref}
\usepackage{url}
\usepackage{bm}
\usepackage{mathrsfs}
\usepackage{dcolumn}
\usepackage{booktabs}
\usepackage{multirow}
\usepackage{stmaryrd}
\usepackage{amsthm}
\usepackage{latexsym}

\usepackage{babel}
\usepackage{lineno}
\linenumbers

\begin{document}

\title{Enhanced Algorithmic Perfect State Transfer on IBM Quantum Computers}

\author{Zong-Yuan Ge$^{1,2}$, Lian-Ao Wu$^{2,3,4}$, Zhao-Ming Wang$^{1,5,6}$}
\thanks{Correspondence and requests for materials should be addressed to Zhao-Ming Wang (email: wangzhaoming@ouc.edu.cn).}
\address{$^1$College of Physics and Optoelectronic Engineering, Ocean University of China, Qingdao 266100, China}
\address{$^2$Department of Physics, University of the Basque Country UPV/EHU, 48080 Bilbao, Spain}
\address{$^3$IKERBASQUE, Basque Foundation for Science, 48013 Bilbao, Spain}
\address{$^4$EHU Quantum Center, University of the Basque Country UPV/EHU, Leioa, 48940 Biscay, Spain}
\address{$^5$Engineering Research Center of Advanced Marine Physical Instruments and Equipment of Ministry of Education, Qingdao 266100, China}
\address{$^6$Qingdao Key Laboratory of Optics and Optoelectronics, Qingdao 266100, China}

\begin{abstract}
Perfect state transfer (PST) through a spin chain can be theoretically obtained via predesigned PST couplings. However, the corresponding experiment on IBM quantum computers demonstrates low transmission success probability (SP) due to noises. Using few qubits of their 127-qubit Eagle processors, we perform the simulation of algorithmic PST through an XY spin chain with PST couplings on \texttt{ibm\_sherbrooke} and \texttt{ibm\_brisbane} processors, alongside Qiskit simulations. The peak SP cannot reach 1 ($\sim0.725$ peak SP for $N=4$). We then propose a comprehensive noise model including Pauli errors, thermal relaxation ($T_1$) and dephasing ($T_2$), and ZZ crosstalk. Based on the experimental parameters provided by the IBM superconducting quantum computing platform, we perform the Qiskit simulation with the comprehensive noise model, and find that the time evolution of the SP is highly consistent with the experimental results. This simulation yields a peak SP of 0.761 at $t \approx \pi/4$, closely matching the results on hardware. To mitigate the impact of noise, we use rescaling techniques to correct noise-induced time shifts and SP decay, achieving an SP improvement of 0.210 (27.60\%) in simulators and 0.263 (38.23\%) on hardware, aligning hitting times closer to ideal values. Additionally, optimal couplings designed via grid search and refined by Bayesian optimization under the comprehensive noise model achieve an SP improvement of 0.190 (26.21\%) in simulators and 0.056 (7.72\%) on hardware. Our work highlights challenges in implementing algorithmic PST on current quantum computers, proposes a comprehensive noise model to effectively describe the system dynamics, and provides insights for developing noise-robust quantum communication protocols.
\end{abstract}

\maketitle

\linenumbers

\section*{Introduction}

Quantum computers, with their quantum hardware and algorithms, have been able to tackle certain complex problems more efficiently than their classical counterparts \cite{Nielsen2010,Ladd}. The well-known Shor algorithm \cite{Shor} provides a paradigmatic example for experimental improvement of computational speed in prime factorizations. Since then, quantum computers have attracted significant global interest and made rapid progress. Specifically, quantum computers demonstrate particular strengths in two domains: simulating physical system behaviors and recognizing informational patterns. The former includes quantum chemistry \cite{Cao2019}, material science \cite{Leon}, and the latter includes quantum biology \cite{Marx} and quantum finance \cite{Herman}, etc. Though it might be hard to fully solve these problems, for the current noisy intermediate-scale quantum (NISQ) devices, the processor scales have expanded to the hundred-qubit level and have shown practical quantum advantages \cite{Preskill2018}. Certain platforms (e.g., IBM Quantum, Quantinuum) are capable of stably executing shallow circuits with depths exceeding one hundred, thereby providing a valuable experimental platform for validating quantum algorithm prototypes \cite{Kandala2019}, molecular ground state simulation \cite{Nam}, and demonstration of quantum-digital payments \cite{schiansky2023demonstration}.

Simulation of quantum system on a classical computer is a hard task, especially for large systems. In contrast, quantum simulation can address this problem by using a controllable quantum system to study another less controllable or accessible system \cite{Nielsen2010,Georgescu2014}. It has been shown that any quantum system evolving under a local Hamiltonian can be efficiently simulated on a quantum computer \cite{Wuetal}. Arbitrary Hamiltonians can be simulated by constructing effective Hamiltonians using Trotter approximation, such as simulation of quantum many-body system dynamics \cite{Smith2019}, dynamically generated decoherence-free subspaces and subsystems \cite{WuReport}, and chemical molecular simulation \cite{Guzik}.

High-fidelity quantum state transfer (QST) is always required in performing quantum information processing tasks \cite{YunlanJi,YunlanJinjp}. For short-distance QST in solid-state quantum systems, spin chains as candidates have been thoroughly investigated \cite{Bose2003}. Normally, for a uniform chain, the transmission fidelity decreases with increasing chain length due to dispersion \cite{Bose2003}. Theoretically, PST through an XY chain can be achieved at a certain time when the coupling strengths are engineered to satisfy mirror symmetry conditions (PST couplings) \cite{Karbach2005,Christandl2004}. Further research demonstrates that a class of one-dimensional XY-type models can support high-fidelity QST, independent of specific coupling configurations \cite{wang2013fault}. Experimentally, QST with PST couplings has been investigated on different platforms, including nuclear magnetic resonance \cite{zhang2005simulation}, photonic lattices \cite{bellec2012faithful,perez2013coherent,chapman2016experimental}, photonic qubit \cite{chapman2016experimental}, and superconducting quantum circuit \cite{xiang2024enhanced,cai2024protecting,zhang2024mech}.

IBM provides an online superconducting quantum computing platform, where few qubits are available freely to the public \cite{IBMQ}. In this paper, within the Qiskit framework, we construct and deploy a quantum circuit based on the Suzuki-Trotter decomposition across multiple IBM Quantum backends. We simulate the algorithmic PST protocols on IBM’s superconducting quantum computing platform, both on hardware and Qiskit simulators.

Quantum computing devices always suffer from environmental noise and system errors, which cause miscalibrated gates, projection noises, measurement errors, etc. As expected, PST on quantum computers cannot be obtained ($\sim0.725$ peak SP for $N=4$) even for the PST couplings. At the same time, PST is highly dependent on system coherence and is particularly sensitive to hardware-induced noise, which makes it a valuable probe for characterizing the noise properties and operational reliability of quantum computing devices. To address the discrepancy between theoretical prediction and practical performance on the IBM quantum computer, we propose a comprehensive noise model based on the experimental parameters provided by the platform, and the two SP evolution curves are highly consistent. We then use two strategies to mitigate the impact of noise on the system based on the comprehensive noise model. The first is quantum error mitigation, where rescaling techniques have been used to enhance the SP by post-processing outputs from an ensemble of circuit runs, achieving an SP improvement of 0.210 (27.60\%) in simulators and 0.263 (38.23\%) on hardware, aligning hitting times closer to ideal values. The second strategy is to optimize the couplings to mitigate the noise impact. We use grid search and Bayesian optimization to search for optimized couplings based on our comprehensive noise model, achieving an SP improvement of 0.190 (26.21\%) in simulators and 0.056 (7.72\%) on hardware. Our work proves that comprehensive noise models are not only feasible but represent an essential step toward approximating real quantum hardware. Although computationally expensive, they can provide indispensable guidance for designing error-correction strategies, optimizing gates, or upgrading hardware.

\section*{Results}

\subsection*{Quantum Chain Configuration}
We have simulated the PST process along a one-dimensional Heisenberg XY chain on the quantum computing platform. The Hamiltonian is
\begin{equation}
H = \sum_{i=1}^{N-1} J_{i,i+1} (\sigma_i^x  \sigma_{i+1}^x  + \sigma_i^y  \sigma_{i+1}^y ), \label{eq:Hamiltonian}
\end{equation}
where $\sigma_i^x $ and $\sigma_i^y $ are Pauli operators for the $i$-th qubit, and $J_{i,i+1}$ is the coupling strength between neighboring qubits. When the couplings satisfy
\begin{equation}
J_{i,i+1} = J_0 \cdot \sqrt{i (N - i)}, \label{eq:Coupling}
\end{equation}
it ensures an equally spaced energy spectrum, and guarantees PST through a chain at time $t = \frac{n\pi}{2}$ with odd $n$ \cite{Salimi2013}. The first hitting time for obtaining the first peak is defined as $t^* = \frac{\pi}{2}$. Here $J_0 = 1$ is the normalized reference coupling.

For simplicity, the initial state is prepared as a single excitation located at the first site of the chain, 
$|\textbf{1}\rangle$=$|10\cdots0\rangle$. The goal is to transfer the excitation to the last site $N$, $|\textbf{N}\rangle$=$|0\cdots01\rangle$, at the designated evolution time. We compute the SP by measuring the excitation probability of the last qubit, defined as
\begin{equation}
P(t) = |\langle \textbf{N} | \psi(t) \rangle|^2, \label{eq:SP}
\end{equation}
where $\psi(t) = U(t) \psi(0)$, $U(t) = e^{-i H t}$ is the system's evolution operator. This transmission mechanism, confined to the single-excitation subspace, has been extensively adopted in theoretical PST studies \cite{christandl2005perfect} and is readily constructible and measurable on real quantum processor \cite{Yung2005}. Note we mainly discuss the transmission of single excitation $|1\rangle$ in the construction of the comprehensive noise model,  the transmission of arbitrary state $A  |0\rangle+B |1\rangle$ will be discussed later. We find that the comprehensive noise model can also be effectively applied to the transmission of arbitrary state.

\subsection*{Simulation and Experimental Setup}
To systematically evaluate the performance of PST in the XY model under ideal and noise conditions, we have constructed a circuit based on Suzuki--Trotter decomposition, with each step implemented using a combination of RXX and RYY gates to realize XY interactions. Comparative experimental analyses have been conducted using the Qiskit framework on a noiseless simulator, a noisy simulator with noise models, and real IBM superconducting processors.

\subsubsection*{Simulation Platforms and Tools}
In this study, Qiskit version 0.45.1 is employed as the primary simulation platform. Two backends provided by the Aer module are utilized: \texttt{aer\_simulator\_statevector} for simulating ideal (noise-free) quantum evolution, and \texttt{qasm\_simulator} combined with a customized \texttt{NoiseModel} for simulating realistic noise effects \cite{QiskitAer}. To simulate the impact of noise on the PST process, we consider three types of representative noises: Pauli channel errors and depolarizing noise, thermal relaxation ($T_1$) and dephasing ($T_2$), ZZ crosstalk. These noise models are commonly employed to characterize non-ideal evolution in quantum computing devices \cite{Geller2013,Cheng2021,Temme2017}. Furthermore, we propose a comprehensive noise model constructed by superimposing multiple error sources, aiming to systematically analyze the SP degradation mechanisms.

\subsubsection*{Experimental Platforms and Equipment Backends}
The real quantum hardware utilized in this study comprises two 127-qubit superconducting processors based on IBM’s Eagle architecture, namely \texttt{ibm\_sherbrooke} and \texttt{ibm\_brisbane}. The latest calibration data---including gate error rates, coherence times, and readout errors---are publicly available on the official IBM Quantum platform \cite{IBMQ}. All experimental circuits are constructed in accordance with the native hardware topology, and the transpilation process is employed to perform gate decomposition and logical qubit mapping tailored to the backend’s specific gate set and connectivity constraints.

Real-device executions are performed via the \texttt{EstimatorV2} interface under the Qiskit Runtime framework. All circuits are transpiled using the \texttt{generate\_preset\_pass\_manager} function with an optimization level set to 3 (\texttt{optimization\_level = 3}), aiming to reduce gate depth and minimize cumulative error during execution \cite{QiskitRuntime}. The final measurement is performed by evaluating the Z-basis expectation value of the qubit at the end of the chain. The SP at each time step is calculated using the following formula
\begin{equation}
P(t) = \frac{1 - \langle \sigma_i^z \rangle}{2}, \label{eq:SP-Z}
\end{equation}
where $i$ denotes the $i$th sites of the chain.

\subsection*{Algorithmic Perfect Quantum State Transfer Simulations}

\subsubsection*{Results on Ideal Simulator and  Real Device }
\begin{figure}[!htb]
    \centering
    \includegraphics[width=1\columnwidth]{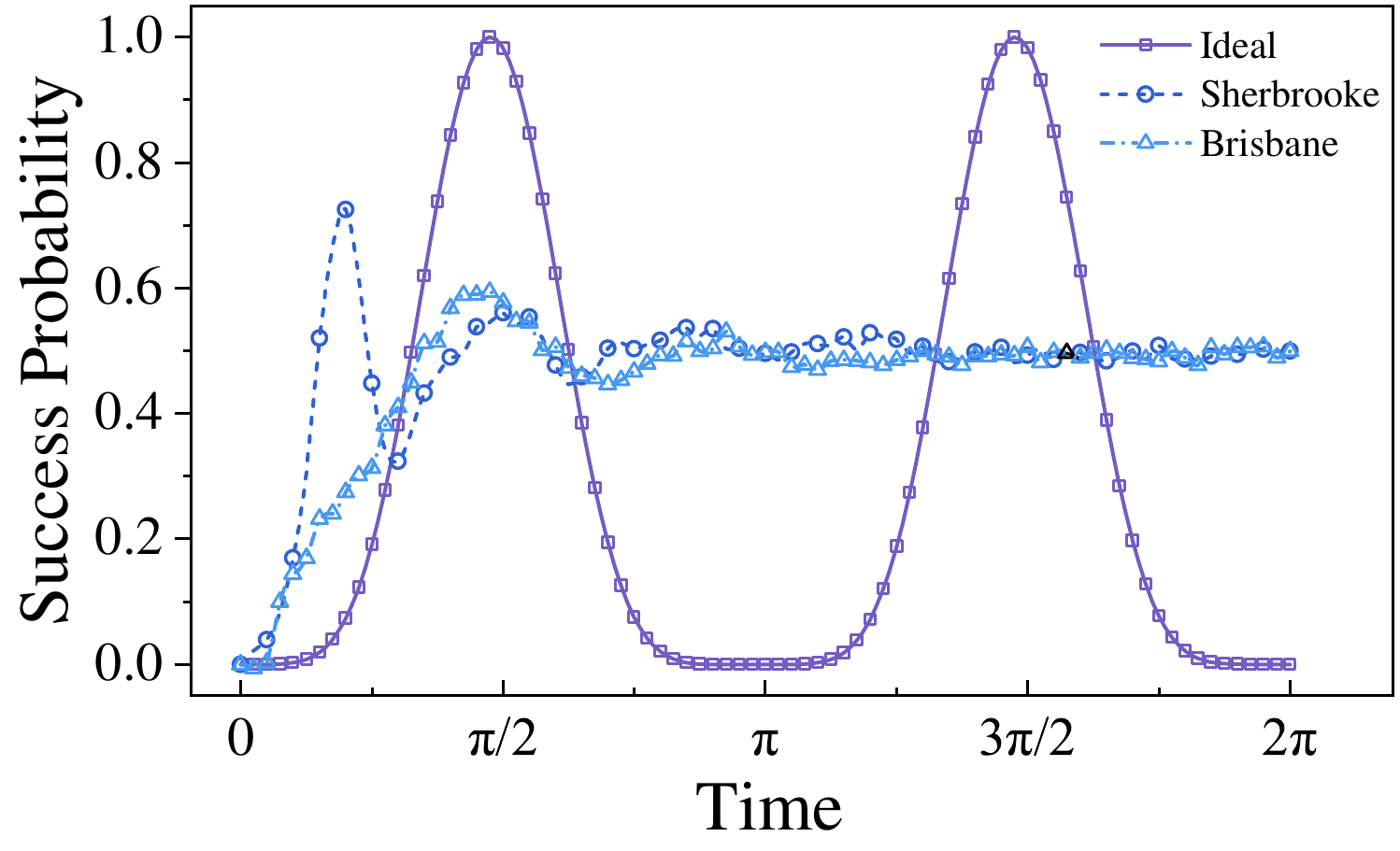}
    \caption{The transmission SP versus time on ideal simulator, \texttt{ibm\_sherbrooke}, and \texttt{ibm\_brisbane} ($N=4$).}
    \label{fig:Ideal-Simulation-noise}
\end{figure}

To investigate the algorithmic PST in the XY chain with PST couplings, we perform simulations on an ideal quantum simulator and real devices for chain lengths $N=4$, which is shown in Fig.~\ref{fig:Ideal-Simulation-noise}. Using the first-order Suzuki-Trotter expansion, we divide the total evolution time $T=2\pi$ into 80 Trotter steps to accurately approximate the time evolution operator. For the ideal simulation, the resulting SP curves demonstrate PST and the peak SP occurs at $t^* = \frac{\pi}{2}$ and $t = \frac{3\pi}{2}$.

For the simulation on real devices, we have implemented the XY chain circuit on IBM Quantum superconducting processors, \texttt{ibm\_sherbrooke} and \texttt{ibm\_brisbane}. We execute the circuit using the Qiskit Runtime framework via the \texttt{EstimatorV2} interface, estimating SP by measuring the $\langle \sigma^z \rangle$ expectation value of the last qubit. The results, shown in Fig.~\ref{fig:Ideal-Simulation-noise}, illustrate the impact of noise on PST performance compared to the ideal simulations.
Clearly, the peak SP cannot reach $1$ in Fig.~\ref{fig:Ideal-Simulation-noise} due to the noise. \texttt{ibm\_sherbrooke} attains a higher peak SP of 0.725 at $t^* = 0.40\times \frac{\pi}{2}$, significantly outperforming \texttt{ibm\_brisbane}’s $0.596$ at $t^* = 0.95\times \frac{\pi}{2}$. Except for peak SP, the hitting time occurs earlier relative to ideal simulations. These differences are likely attributable to variations in qubit connectivity, coherence times, and gate fidelities between the devices.

Further analysis of hardware parameters reveals that \texttt{ibm\_sherbrooke} has longer coherence times ($T_1 = 266.74 \, \mu s$, $T_2 = 199.97 \, \mu s$) compared to \texttt{ibm\_brisbane} ($T_1 = 242.82 \, \mu s$, $T_2 = 129.75 \, \mu s$). Additionally, \texttt{ibm\_sherbrooke} exhibits a lower two-qubit gate error rate (1.51\% vs. 2.05\% for \texttt{ibm\_brisbane}). These factors enable \texttt{ibm\_sherbrooke} to maintain higher coherence and gate accuracy compared to \texttt{ibm\_brisbane}, contributing to improved performance. Due to \texttt{ibm\_sherbrooke}’s superior coherence and lower error rates, we study $N=4$ chains on this platform to explore noise effects in QST, addressing scaling challenges in later discussions.

At last, we stress that although the results on the IBM quantum computers are quantitatively inconsistent with ideal simulation, they demonstrate clear quantitative behavior: oscillation with time. This indicates that carefully designed circuits and effective noise mitigation strategies \cite{Li2017,Kandala2017} allow current superconducting quantum hardware to approximate the algorithmic PST dynamics observed in simulations.

\subsection*{Quantum Noise Impact Simulation}
To estimate the impact of noise on the transmission SP in real devices, we have conducted a simulation-based analysis of various noise mechanisms, using parameters matched to \texttt{ibm\_sherbrooke}’s hardware characteristics. The simulation includes three primary noise sources: Pauli errors (including depolarization), decoherence ($T_1$/$T_2$), and crosstalk \cite{Nielsen2010}. We first study each noise type's effect independently and elucidate their relative contributions to the SP degradation, then we develop an optimization model to improve SP based on the contributions of the comprehensive noise effects \cite{Babukhin2022}.

To ensure comparability with experimental results, we configure the noise model using \texttt{ibm\_sherbrooke}’s gate error rate (1.51\% for two-qubit gates), readout error, coherence times ($T_1 = 266.74 \, \mu s$, $T_2 = 199.97 \, \mu s$), and qubit connectivity. The simulation employs 80 Trotter steps, consistent with the experimental circuit, with noise parameters adjusted to replicate hardware conditions.

The parameters are configured with reference to the \texttt{ibm\_sherbrooke} characteristics, with the aim of minimizing discrepancies due to error, and are specified as follows

\begin{table}[!htbp]
    \centering
    \caption{Noise model parameters used in simulations, based on \texttt{ibm\_sherbrooke} characteristics.}
    \label{tab:Noise-Parameters}
    \begin{tabular}{llc}
        \toprule
        Type of Noise & Parameter Settings & Affected Gates \\
        \midrule
        Pauli Noise & $p = 1.875 \times 10^{-3}$ & x, rx, ry, rz \\
        Depolarizing Noise & $q = 2.5 \times 10^{-3}$ & All gates \\
        Thermal Relaxation & $T_1 =  266.74 \, \mu s$ & All gates \\
        Dephasing & $T_2 = 199.97 \, \mu s$ & All gates \\
        Crosstalk (ZZ) & $\zeta = 0 \sim 0.3 \, \mathrm{MHz}$ & ZZ-phase gates \\
        Comprehensive Noise & Parameters above & All gates \\
        \bottomrule
    \end{tabular}
\end{table}

\subsubsection*{Pauli and Depolarizing Noise}
To assess the impact of local Pauli-type noise, we select Pauli noise and depolarizing noise as the focus of this section.For single-qubit operations, the Pauli noise is represented by the quantum channel,
\begin{equation}
\mathcal{E}(\rho) = (1 - p) \rho + p_x X \rho X + p_y Y \rho Y + p_z Z \rho Z, \label{eq:Pauli-Noise}
\end{equation}
where $p$ is the total error probability, and $p_x, p_y, p_z$ represent the error probabilities of the X, Y, and Z channels, respectively, satisfying $p_x + p_y + p_z = p$. For the depolarizing noise, these error probabilities are equal ($p_x = p_y = p_z = \frac{p}{3}$) \cite{Flammia2020,Chen2023}. Notably, depolarizing noise is a special case of Pauli noise \cite{Urbanek2021}, commonly used as a benchmark for quantum gate errors. In quantum computing, depolarizing noise describes the degradation of a qubit to a fully mixed state,
\begin{equation}
\mathcal{E}_{\mathrm{depol}}(\rho) = \left(1 - \frac{3}{4} q\right) \rho + \frac{q}{4} (X \rho X + Y \rho Y + Z \rho Z). \label{eq:Depolarizing-Noise}
\end{equation}
The relationship between Pauli and depolarizing error probabilities is given by $q = 4p/3$.

For two-qubit gates, Pauli noise is typically modeled as the tensor product of two single-qubit noise channels, assuming uncorrelated errors between qubits. The two-qubit Pauli noise channel is expressed as
\begin{equation}
\mathcal{E}_{2q}(\rho) = \mathcal{E}_1(\rho) \otimes \mathcal{E}_2(\rho), \label{eq:Two-Qubit-Pauli}
\end{equation}
where $\mathcal{E}_i(\rho)$ $(i=1,2)$ denotes the single-qubit Pauli noise channel.

All parameters are set with reference to the hardware parameters of the \texttt{ibm\_sherbrooke} device on the day of the experimental run, including its median single-qubit gate error ($\approx 2 \times 10^{-4}$) and two-qubit gate error ($\approx 6.3 \times 10^{-3}$) \cite{IBMQ}. Each Trotter step in this simulation consists of six two-qubit gates, with the noise channel applied only to each target gate. Thus, $p$ represents the effective error after gate stacking. Since the effect of $T_1$, $T_2$ decoherence is readily quantified, we found that it degrades SP by approximately 15\%. Additionally, ZZ crosstalk has a minimal impact on SP. From the experimental results, we deduce that the total cumulative error from Pauli noise in the simulation must be maintained within a reasonable range of 15\% to 20\%.

\begin{figure}[!htb]
    \centering
    \subfigure{\includegraphics[width=1\columnwidth]{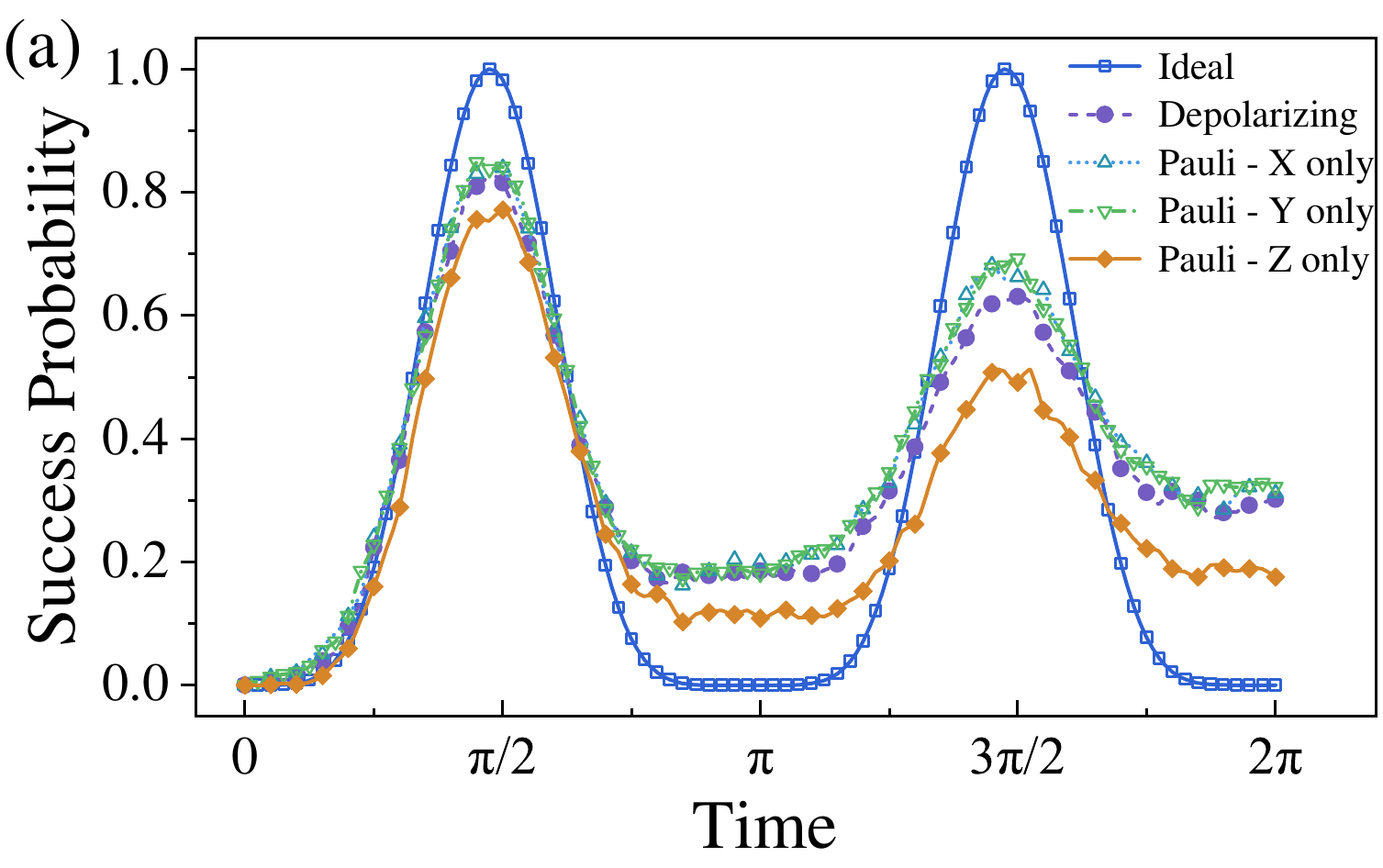}}
    \subfigure{\includegraphics[width=1\columnwidth]{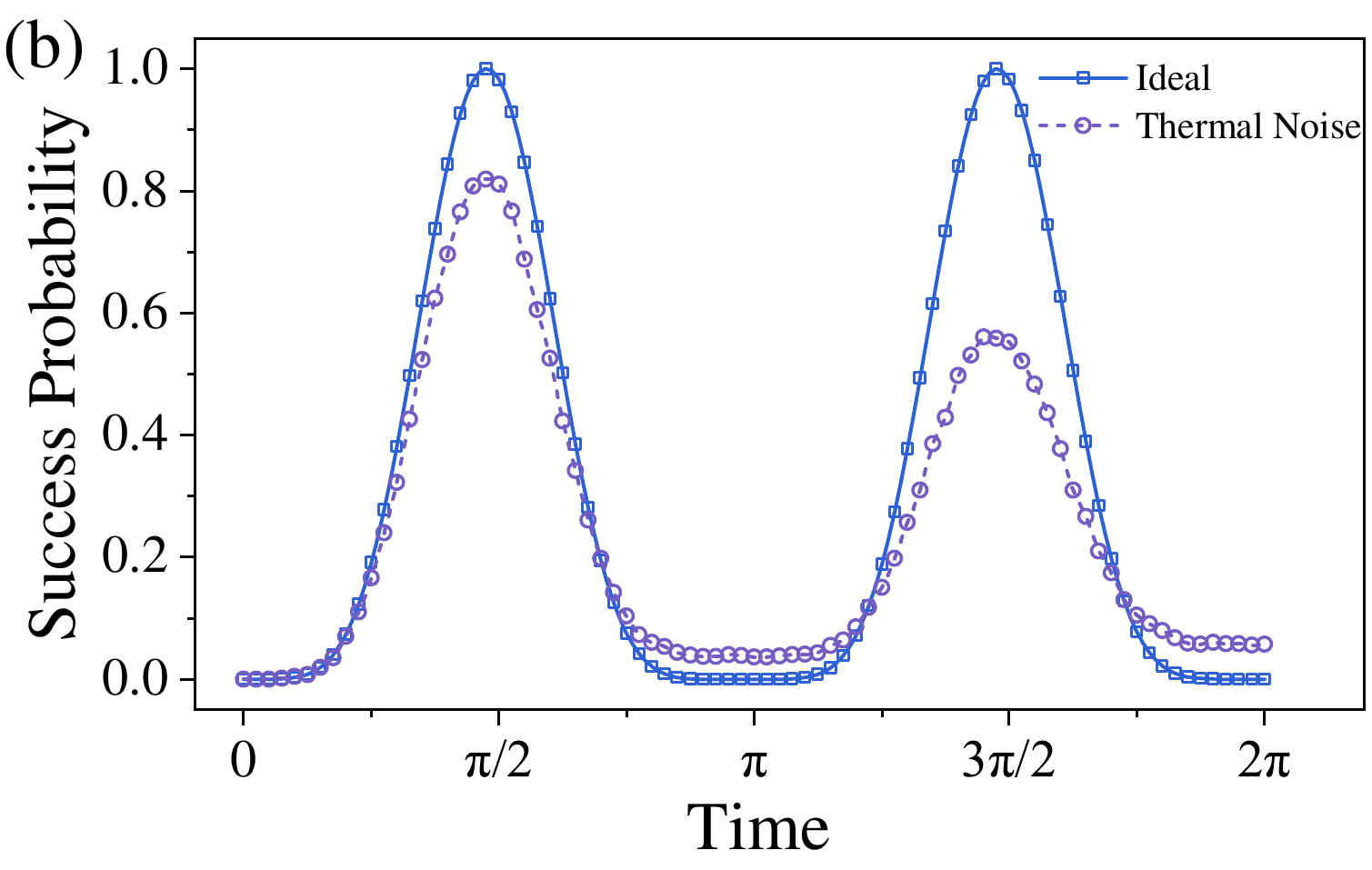}}
    \subfigure{\includegraphics[width=1\columnwidth]{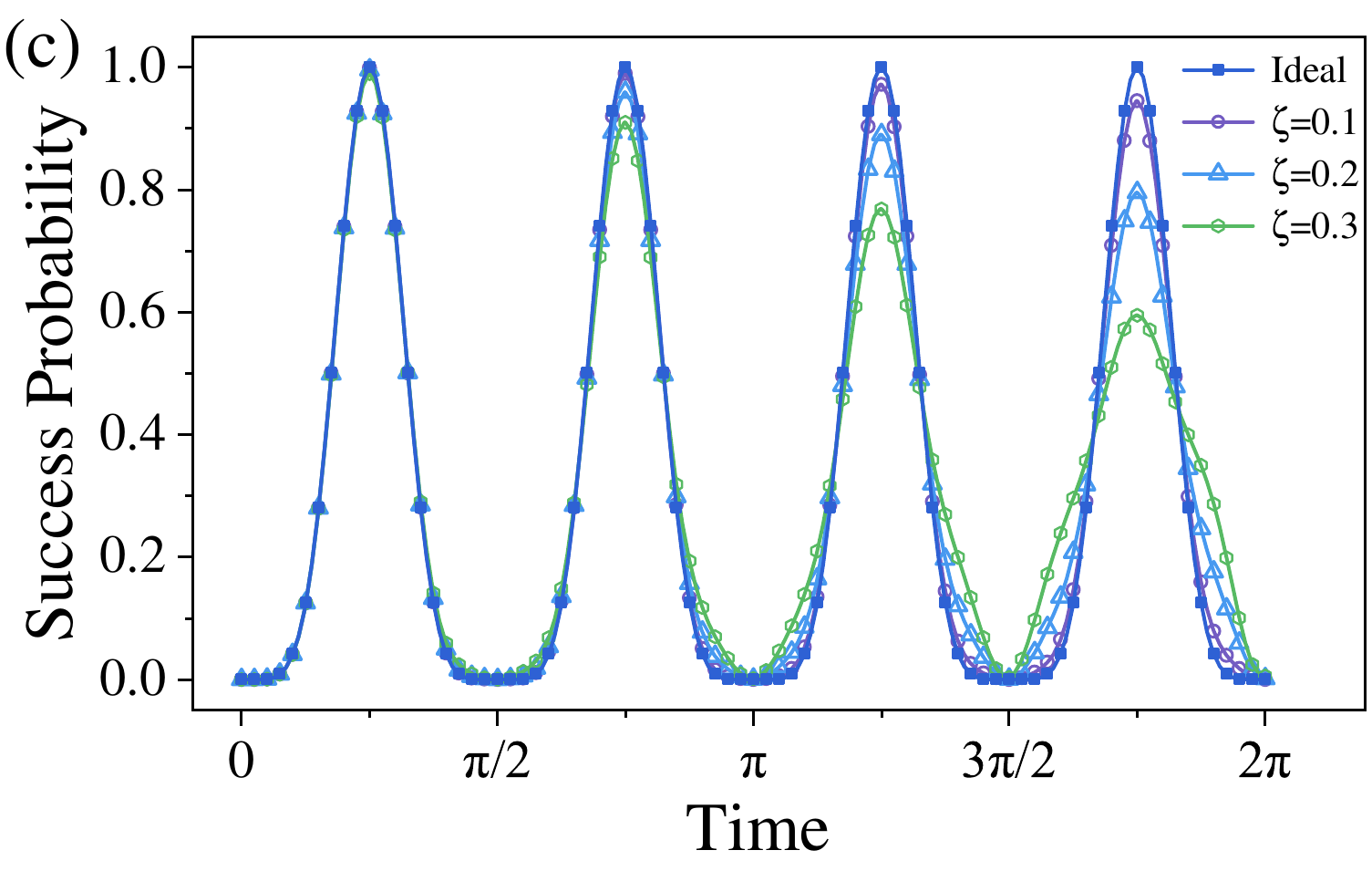}}
    \caption{Time evolution of the SP under different types of noise for PST simulation (\( N=4 \)). (a) Pauli and Depolarizing Noise; (b) \( T_1/T_2 \) Decoherence; (c) ZZ Crosstalk; The noise model parameters are taken from Table~\ref{tab:Noise-Parameters}, which are configured with reference to the \texttt{ibm\_sherbrooke} device on the day of the experimental run.}
    \label{fig:Noise}
\end{figure}

By comparing SP evolution under different types of noise, in Figure~\ref{fig:Noise} (a) we observe distinct effects of Pauli noise channels on QST. Furthermore, the depolarizing channel, serving as a baseline model, exhibits uniform noise, leading to a smoother QST process with minimal dynamic perturbations compared to Pauli noise channels \cite{Escofet2025,Gonzalez-Garcia2025}. We observe that different noise models significantly alter system SP, consistent with theoretical predictions. Among the noise models, the Pauli Z channel has the strongest impact on SP degradation due to its disruption of phase coherence. The depolarizing channel follows with a moderate effect, while the Pauli X and Y channels exhibit similar and relatively weaker influences.

\subsubsection*{\texorpdfstring{$T_1$/$T_2$}{T1/T2} Decoherence}
In real quantum devices, $T_1$ relaxation and $T_2$ dephasing always exist and they significantly decrease the SP. Study of these effects is crucial for improving the quantum information processing \cite{Smith2019,Geller2013}. $T_1$ relaxation for a single qubit is modeled by
\begin{equation}
\mathcal{E}_{T_1}(\rho) = (1 - \gamma_1) \rho + \gamma_1 |0\rangle\langle0|, \label{eq:T1-Relaxation}
\end{equation}
where $\gamma_1 = 1 - e^{-t/T_1}$ is the relaxation factor, describing energy decay over $T_1$. $T_2$ represents the dephasing time
\begin{equation}
\mathcal{E}_{T_2}(\rho) = \gamma_2 \rho + (1 - \gamma_2) Z \rho Z, \label{eq:T2-Dephasing}
\end{equation}
where $\gamma_2 = e^{-t/T_2}$ is the decoherence factor, describing the loss of qubit phase coherence. For two-qubit gate operations, $T_1$ and $T_2$ decoherence affect qubit pairs via a tensor product
\begin{equation}
\mathcal{E}_{2q}(\rho) = \mathcal{E}_{T_1/T_2}(\rho) \otimes \mathcal{E}_{T_1/T_2}(\rho). \label{eq:Two-Qubit-T1T2}
\end{equation}

\paragraph*{\texorpdfstring{$T_1$, $T_2$}{T1, T2} Parameter Settings}
In this study, we set $T_1 = 266.74 \, \mu s$ and $T_2 = 199.97 \, \mu s$ based on the hardware parameters of the \texttt{ibm\_sherbrooke} device. To simulate $T_1$ relaxation and $T_2$ dephasing accurately, we use Qiskit’s \texttt{thermal\_relaxation\_error} function, which incorporates thermal relaxation and dephasing errors into each quantum gate operation, isolating $T_2^*$ to avoid redundant modeling of $T_1$ contributions.

\paragraph*{Gate Time Settings}
The duration of single-qubit and two-qubit gates is a key factor affecting the SP of quantum operations. In this simulation, we set the single-qubit gate duration to 57 ns, reflecting the time required for single-qubit operations and influencing their SP \cite{Brown2024}. Two-qubit gates are more complex, as they involve interactions between qubits, resulting in longer operation times \cite{McKay2023}. For two-qubit gates (e.g., CNOT and ECR gates), we set an operation duration of 533 ns.

The impact of $T_1$ and $T_2$ decoherence is significant, as shown in Figure \ref{fig:Noise}(b). The SP degradation is primarily driven by $T_2$ dephasing for short chains, with $T_1$ relaxation becoming more pronounced as the chain length increases \cite{Magesan2020,Mitchell2021}.

\subsubsection*{ZZ Crosstalk}
ZZ crosstalk is induced by unwanted couplings between qubits in superconducting quantum processors, particularly in architectures like \texttt{ibm\_sherbrooke}. It manifests as a residual $\sigma_z \otimes \sigma_z$ interaction between neighboring qubits, causing unwanted phase accumulation and frequency shifts \cite{Magesan2020,Mitchell2021}.

For fixed-frequency qubits, ZZ crosstalk is modeled by the quantum channel
\begin{equation}
\Phi(\rho) = U_{zz} \rho U_{zz}^\dagger, \label{eq:ZZ-Crosstalk}
\end{equation}
where $U_{zz}$ is the unitary operator for the ZZ interaction, expressed as
\begin{equation}
U_{zz}(t) = \begin{pmatrix}
e^{-i \zeta t} & 0 & 0 & 0 \\
0 & e^{i \zeta t} & 0 & 0 \\
0 & 0 & e^{i \zeta t} & 0 \\
0 & 0 & 0 & e^{-i \zeta t}
\end{pmatrix},\label{eq:ZZ-Unitary}
\end{equation}
with $\zeta$ as the ZZ interaction constant. This model describes the ZZ interaction between neighboring qubits, which intensifies with increasing circuit execution time. Thus, ZZ crosstalk is particularly significant in moderately deep circuits on quantum processors.

To investigate the effects of ZZ crosstalk on QST, we adjust the interaction constant $\zeta$ in numerical simulations. We set $\zeta \in \{0.0, 0.05, 0.1, 0.2\} \, \mathrm{MHz}$, a range covering typical ZZ coupling strengths in current superconducting quantum processors and their potential extremes \cite{Ni2022}. Figure~\ref{fig:Noise}(c) shows that simulations of ZZ crosstalk noise have a relatively small impact on SP. Particularly in shorter quantum circuits, ZZ crosstalk noise causes minimal reduction in the peak SP. The primary effect of ZZ crosstalk noise is observed in altered transfer times, manifesting in PST as an earlier hitting time and a shortened period \cite{Ni2022}.

\subsubsection*{Treating ZZ crosstalk as an effect Hamiltonian}
By incorporating ZZ crosstalk as a Hamiltonian $H_{zz} = \zeta \sigma_z \otimes \sigma_z$ ($\zeta = 0.1 \, \text{MHz}$) into the system evolution \cite{Magesan2020}, our model precisely captures its coherent impact on the hitting time of PST, significantly outperforming traditional random noise channel methods \cite{Krinner2022,Tripathi2022}. Traditional approaches typically model ZZ crosstalk as a random dephasing channel, expressed as
\begin{equation}
\mathcal{E}_{\text{zz}}(\rho) = (1 - p_{\text{zz}}) \rho + p_{\text{zz}} (\sigma_z \otimes \sigma_z) \rho (\sigma_z \otimes \sigma_z).
\end{equation}
This assumes ZZ crosstalk as incoherent noise, applying discrete phase flips with probability $p_{\text{zz}}$ only after gate operations. Such method neglects the persistent coherent nature of parasitic coupling in superconducting quantum processors, failing to accurately predict the systematic shift in the hitting time, and instead merely results in broadened or reduced SP peaks \cite{Krinner2022}. In contrast, our approach simulates the continuous phase evolution of ZZ crosstalk using \texttt{rzz} gates (rotation angle $2 \zeta \Delta t$) within the total Hamiltonian $H = H_{\text{XY}} + H_{zz}$, accurately reflecting its coherent impact on QST paths \cite{Ni2022}.

\subsubsection*{Comprehensive Noise Simulation}

\begin{figure}[!htb]
    \centering
    \includegraphics[width=1\columnwidth]{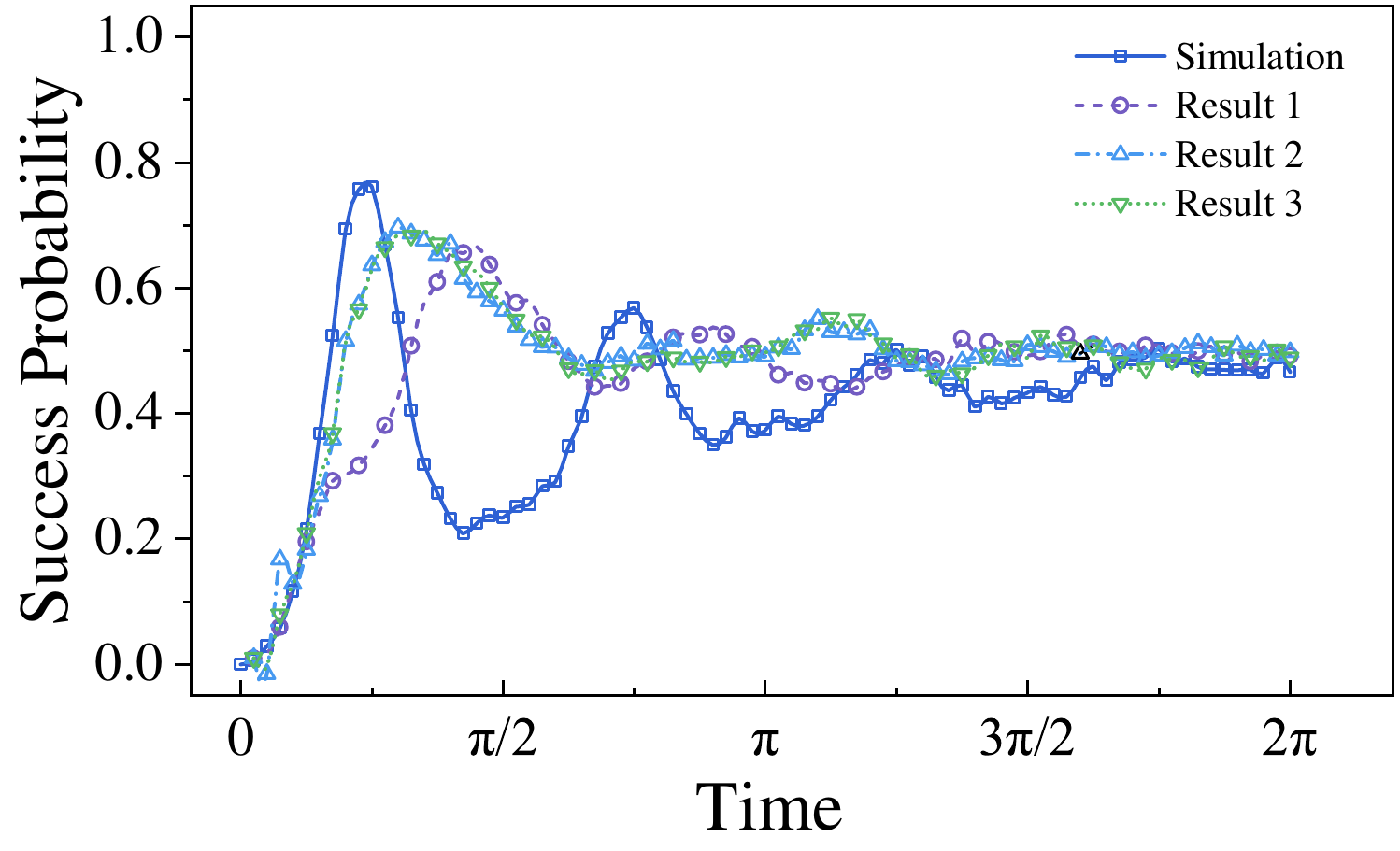}
    \caption{SP evolution under comprehensive noise model compared to multiple experimental runs on \texttt{ibm\_sherbrooke} ($N=4$). The solid line represents the simulated SP, showing a first-period peak of $\sim0.76$ at $t^* \approx \pi/4$, a second-period peak of $\sim0.6$, and fluctuations around 0.5 after $t = \pi$, while dotted line indicates experimental data.}
    \label{fig:Comprehensive-Noise}
\end{figure}

To evaluate the impact of multiple noise sources on algorithm PST, we propose a comprehensive noise model integrating Pauli noise (with depolarizing noise as a special case), decoherence ($T_1$/$T_2$), and ZZ crosstalk. The model is designed to capture the synergistic effects of these noise sources, reflecting the complex interactions observed in real quantum hardware. Noise parameters are calibrated to \texttt{ibm\_sherbrooke} data, with Pauli noise probabilities set to $p_x = p_y = p_z = 1.875 \times 10^{-3}/3$ (equivalent to depolarizing noise with $q = 2.5 \times 10^{-3}$), $T_1 = 266.74 \, \mu\text{s}$, $T_2 = 199.97 \, \mu\text{s}$, and ZZ crosstalk strength $\zeta = 0.1 \, \text{MHz}$ \cite{Ni2022}. From Figure \ref{fig:Comprehensive-Noise}, the first peak SP is about 0.7 for all the experimental runs on \texttt{ibm\_sherbrooke}, i.e., the total observed SP shows 20\%-30\% decrease.  Depolarizing noise models gate errors, contributing 10--15\% to SP loss, while $T_1$/$T_2$ decoherence, applied via Qiskit’s \texttt{thermal\_relaxation\_error} to gates with durations of 57 ns (single-qubit) and 533 ns (two-qubit), accounts for 15--20\% loss. ZZ crosstalk, modeled as a Hamiltonian $H_{zz} = \zeta \sigma_z \otimes \sigma_z$ with \texttt{rzz} gates (rotation angle $2 \zeta \Delta t$), captures coherent phase evolution and hitting time shifts (from $t^* = \pi/2$ to $t^* \approx \pi/4$) with minimal SP impact ($\sim$1\%) \cite{Magesan2020,Krinner2022,Tripathi2022}. 

The design framework prioritizes precise replication of experimental SP degradation via layered noise-source integration. Depolarizing noise is applied first to model gate operation errors, with its probability tuned to match the residual SP loss after subtracting the contributions of $T_1$/$T_2$ decoherence and ZZ crosstalk. Next, $T_1$/$T_2$ decoherence is layered to capture environmental noise affecting qubit coherence, compounding gate errors as environmental decoherence contributes to overall gate infidelity. Finally, ZZ crosstalk is incorporated within each Trotter step to simulate coherent phase shifts, capturing the systematic hitting time shift. 

The simulation, using an 80-step Suzuki-Trotter decomposition over $T = 2\pi$ with initial state $|1000\rangle$ and target state $|0001\rangle$, yields an SP peak of 0.76 at $t^* \approx \pi/4$, matching experimental results on \texttt{ibm\_sherbrooke} (Figure \ref{fig:Comprehensive-Noise}). The sequence of noises ensure the model captures the synergistic interplay of noise sources. For example, increasing $T_1$ and $T_2$ by a factor of 10 in the comprehensive model only improved peak SP by 0.02, far less than the near-complete recovery observed in isolated $T_1$/$T_2$ simulations, demonstrating that the synergistic interactions among noise sources significantly reduce the impact of individual parameter changes compared to single-noise models.

\subsubsection*{Time Evolution of SP on various Sites}
To visualize the dynamic evolution of PST over the spin chain, we measure the SP on each site in experiments, take simulations under ideal conditions, comprehensive noise model, and on \texttt{ibm\_sherbrooke}. The experimental setup is consistent with the prior section, with a total evolution time of $T = 2\pi$, and the Suzuki-Trotter evolution uses 20 steps to minimize noise accumulation and match previous hardware performance.

\begin{figure}[!htb]
    \centering
    \subfigure{\includegraphics[width=0.9\columnwidth]{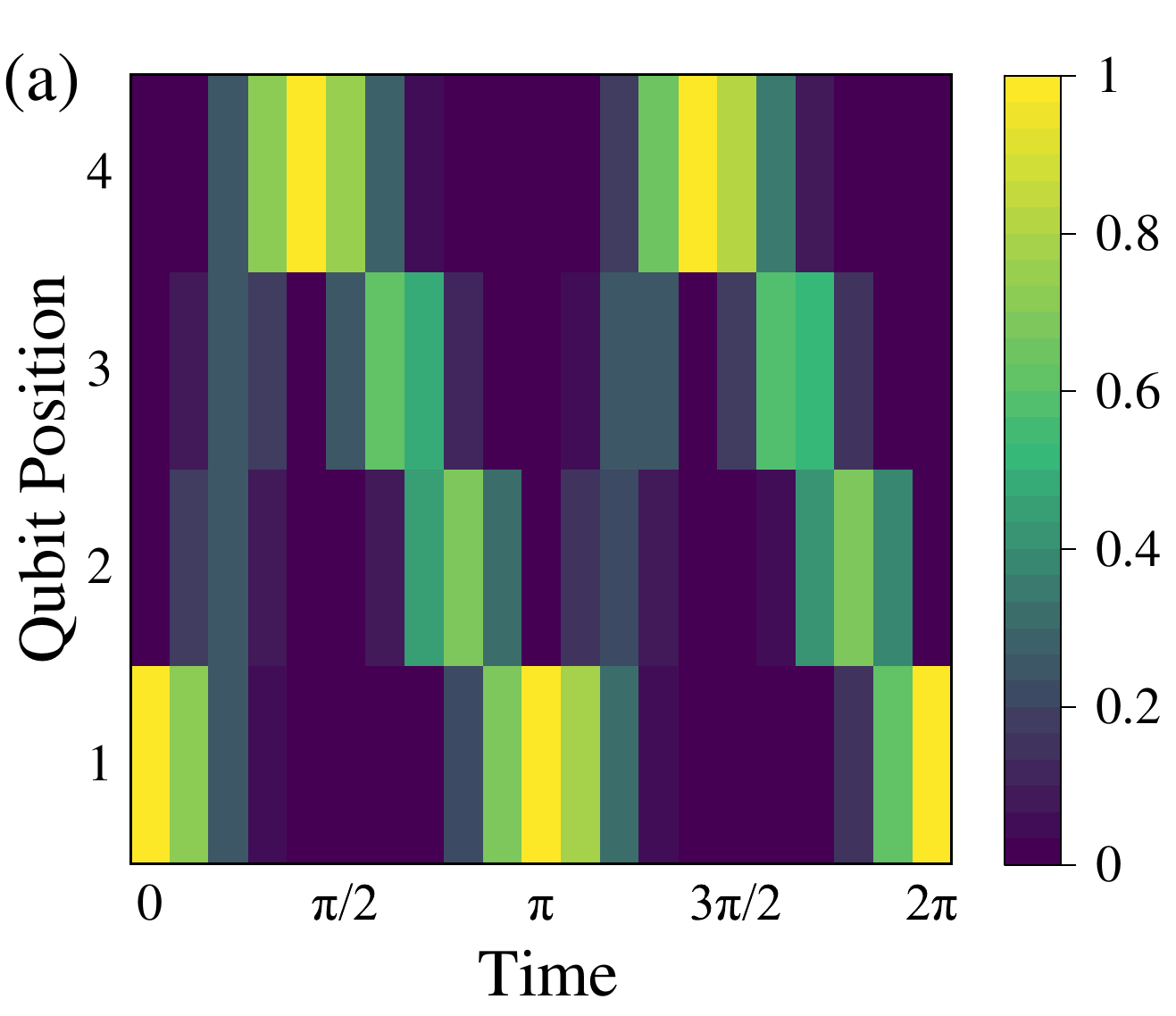}}
    \subfigure{\includegraphics[width=0.9\columnwidth]{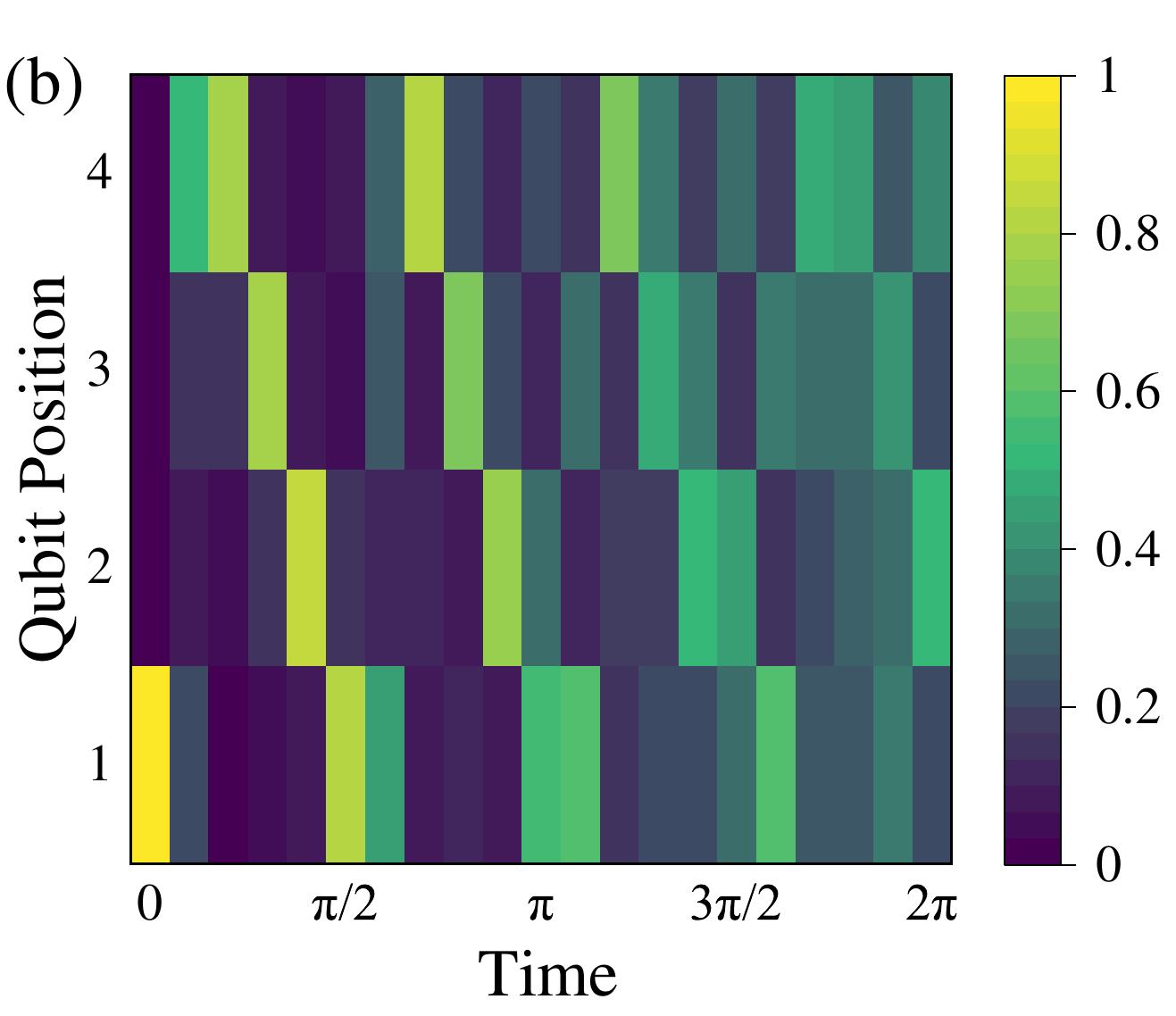}}
    \subfigure{\includegraphics[width=0.9\columnwidth]{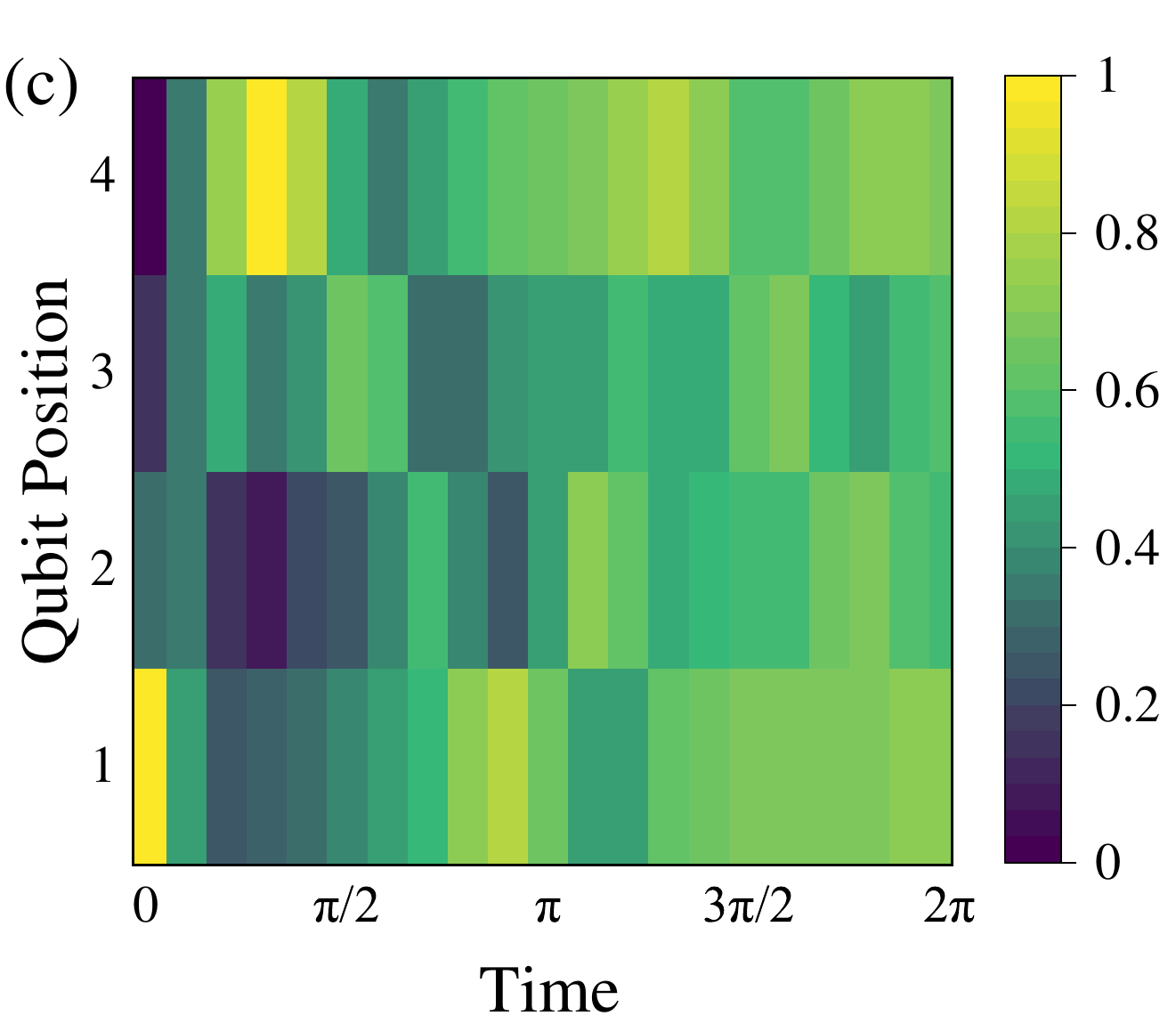}}
    \caption{SP evolution along the sites in (a) ideal PST simulation; (b) simulation under comprehensive noise; (c) on \texttt{ibm\_sherbrooke}.}
    \label{fig:Positions}
\end{figure}

Under ideal conditions, QST exhibits periodic dynamics. At $t=0$, the SP at position 1 is $\text{P}_1(0) = 1$. As time evolves, the excitation propagates from site 1 to site 4, peaking at $\text{P}_4(t^*) = 1$ at $t^* = \frac{\pi}{2}$, with $\text{P}_1$ dropping to zero, indicating PST. By $t = \frac{3\pi}{2}$, the state returns to site 1, mirroring the initial distribution and confirming the expected periodic oscillations.

In the comprehensive noise simulation, QST retains a periodic behavior similar to the ideal simulation, though modified by noise effects, with the peak SP at site 4 reaching only $\text{P}_4 = 0.76$ and the hitting time advancing to $t^* \approx \pi/4$ compared to the ideal $t^* = \pi/2$. By $t \approx \pi$, the periodicity becomes indistinct. The peak SP drops to approximately 0.58, indicating significant decoherence due to multiple noise sources.

On \texttt{ibm\_sherbrooke}, noise significantly affects the results. In the first period, the state transfers from site 1 to position 4 with certain SP, with $\text{P}_4 \approx 0.72$ at $t^* \approx \frac{\pi}{4}$, consistent with prior comprehensive noise simulations. After $t > \pi$, coherence decreases significantly, and the SP distribution becomes heterogeneous, primarily due to cumulative $T_1$/$T_2$ decoherence and depolarizing noise.

\subsection*{Quantum Error Mitigation}

\subsubsection*{Rescaling Technique to Improve SP}
To evaluate the impact of noise on the SP and hitting time of algorithmic PST, we use a rescaling technique that corrects noise-induced time shifts and SP decay to restore ideal dynamic evolution. This technique relies on the comprehensive noise model we proposed before, assuming exponential SP decay with Trotter steps and aligning hitting times via temporal scaling \cite{Babukhin2022}.

The rescaling model first computes a time scaling factor by comparing peak times from ideal and noisy simulations
\begin{equation}
s = \frac{t_{\mathrm{ideal}}}{t_{\mathrm{simulation}}}, \label{eq:Time-Scaling}
\end{equation}
where $t_{\mathrm{ideal}}$ is the ideal hitting time in the first period, and $t_{\mathrm{simulation}}$ is the hitting time in experimental or noisy simulations. The noisy time axis is then adjusted as
\begin{equation}
t_{\mathrm{scaled}} = t_{\mathrm{noise}} \cdot s, \label{eq:Scaled-Time}
\end{equation}
which aligns the noisy evolution with the ideal hitting time. Using the scaled time axis, the corresponding ideal SP $\text{P}_{\mathrm{ideal}}(t_{\mathrm{scaled}})$ is obtained via interpolation.

SP is then corrected by using
\begin{equation}
n(t_{\mathrm{scaled}}) = \frac{\hat{n}(t_{\mathrm{scaled}}) - \alpha (1 - e^{-\beta k})}{e^{-\beta k}}, \label{eq:Rescaling-SP}
\end{equation}
where $k$ is the number of Trotter steps, $\hat{n}(t_{\mathrm{scaled}})$ is the noisy SP, and $n(t_{\mathrm{scaled}})$ is the corrected SP. The parameter $\alpha$ represents the noise-induced SP offset, and $\beta$ is the decay rate.

\begin{figure}[!htb]
    \centering
    \includegraphics[width=1\columnwidth]{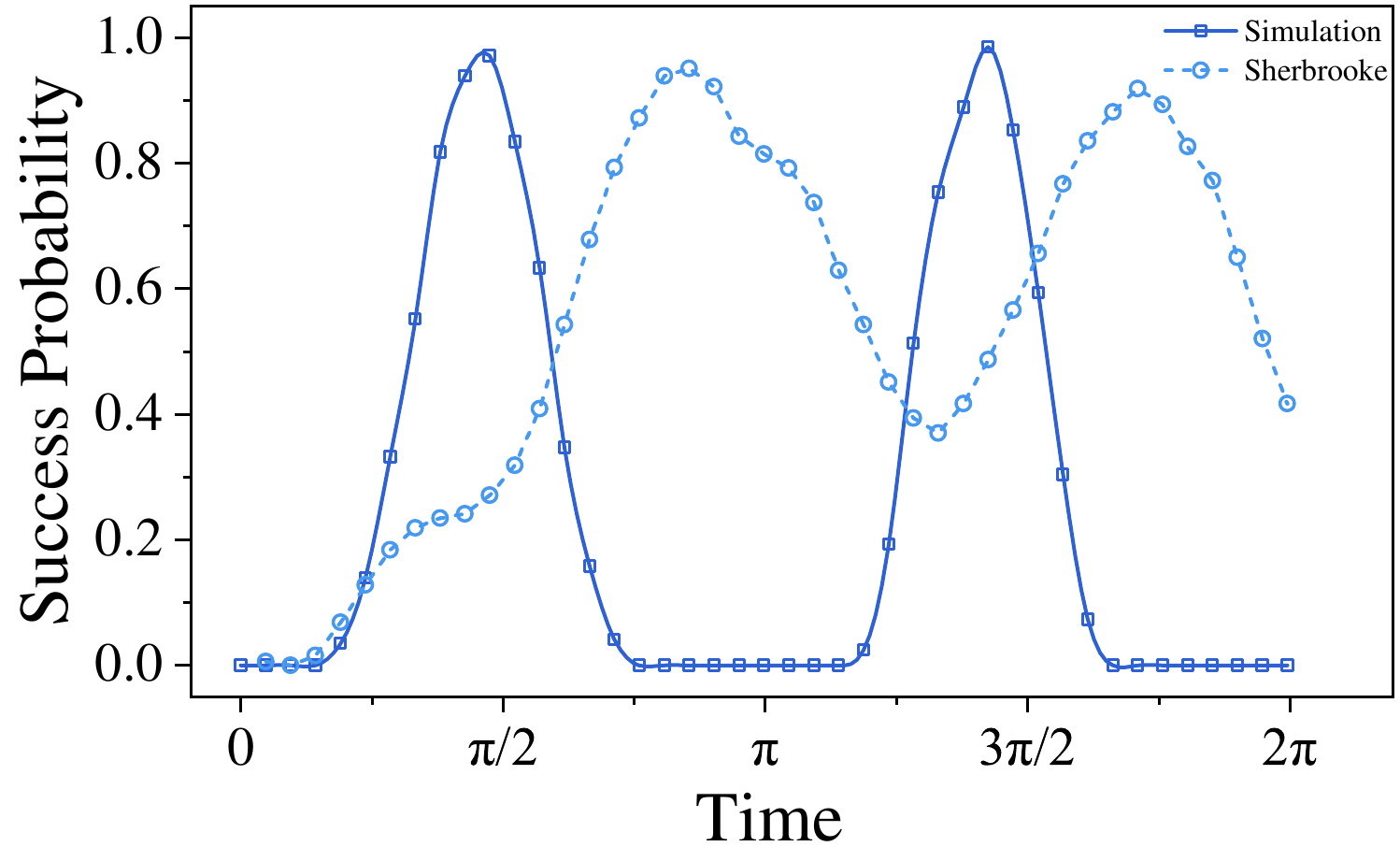}
    \caption{Time evolution of the mitigated SP for two cases: simulations under the comprehensive noise and on \texttt{ibm\_sherbrooke}.}
    \label{fig:Rescaled-Simulation}
\end{figure}

In the comprehensive noise simulation, we obtain $\alpha = 0.463$ and $\beta = 0.054$ from fitting. From Figure~\ref{fig:Rescaled-Simulation}, the uncorrected first peak SP occurs at approximately 0.761, and rescaling increases the peak to 0.971 in the first period and 0.985 in the second period, reducing the error by about 0.210 (27.6\%). This indicates that rescaling effectively mitigates the degradation caused by depolarizing noise while compensating for amplitude decay due to $T_1$/$T_2$ decoherence. The fitted parameters align with theoretical expectations, confirming their role in addressing noise-induced SP degradation.

On \texttt{ibm\_sherbrooke}, from Figure~\ref{fig:Rescaled-Simulation}, the uncorrected first SP peaks at approximately 0.688. Using the simulation-fitted $\alpha$ and $\beta$, rescaling raises the peak to 0.951 in the first period and 0.919 in the second period, reducing the error by about 0.263 (38.23\%). Rescaling effectively corrects the cumulative effects of gate errors and $T_1$/$T_2$ decoherence. Note that the corrected SP is slightly lower than the one in simulations, likely due to non-Markovian noise in the device \cite{Strikis2021}.

\subsubsection*{Optimal Coupling Design via Grid Search and Bayesian Optimization}
\begin{figure}[!htb]
    \centering
    \subfigure{\includegraphics[width=1\columnwidth]{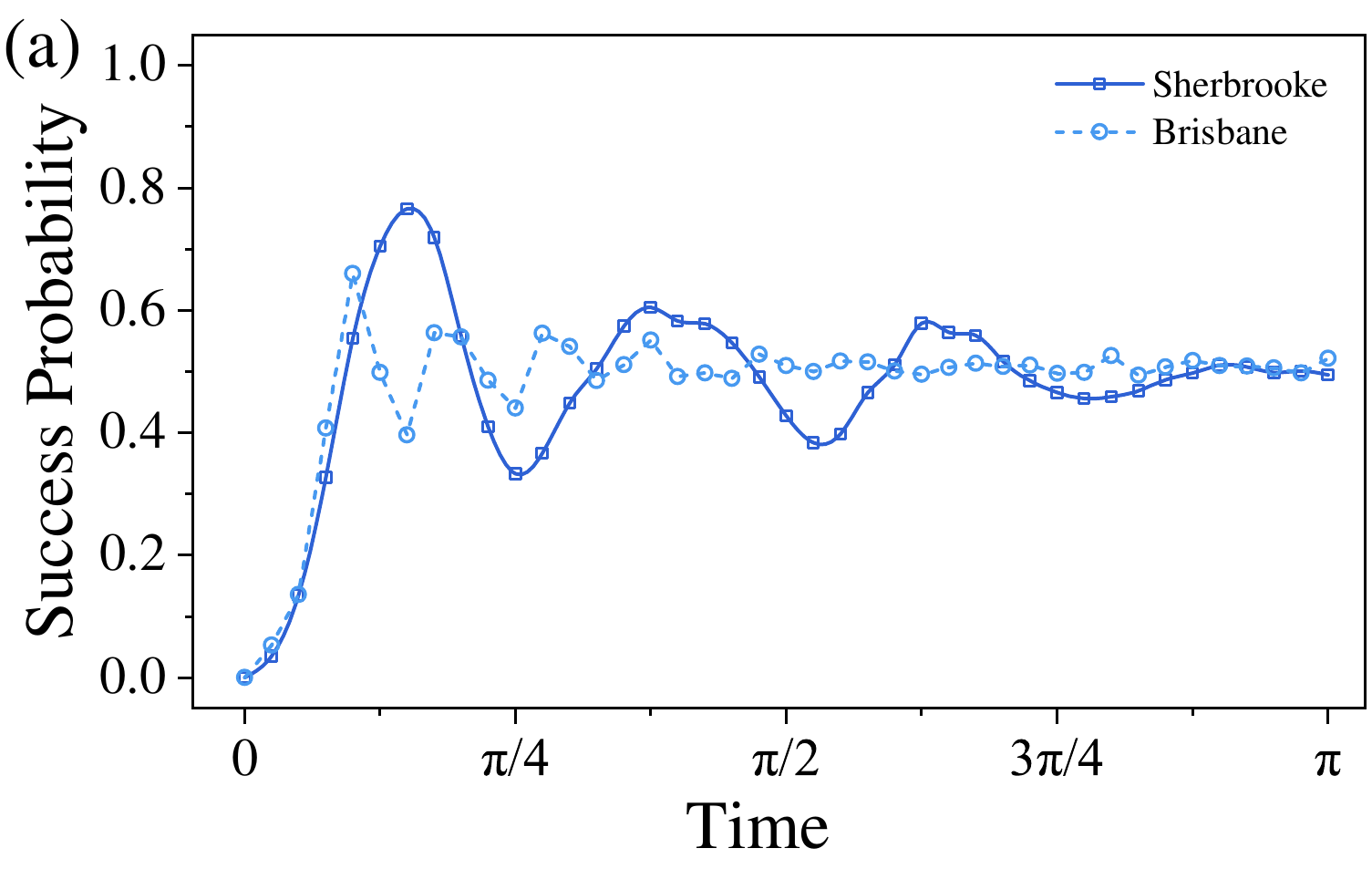}}
    \subfigure{\includegraphics[width=1\columnwidth]{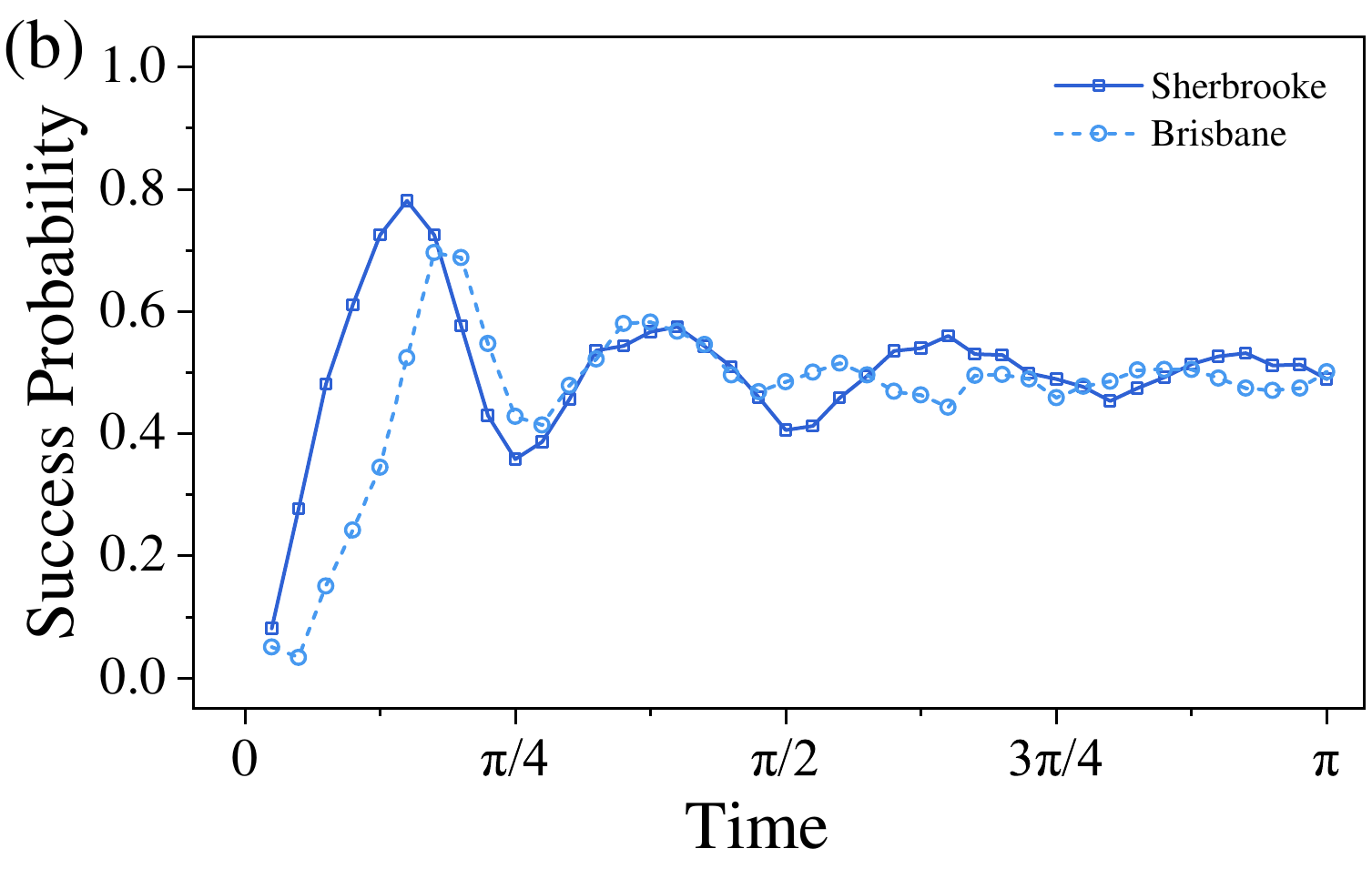}}
    \caption{SP evolution on \texttt{ibm\_sherbrooke} with (a) optimized uniform coupling strengths ($J_0 = 2.9$); (b) optimal non-uniform coupling strengths ($J_{1,2}=2.9788, J_{2,3}=3.0182, J_{3,4}=2.8212$).}
    \label{fig:Grid-Search}
\end{figure}

To enhance the robustness of QST in noisy environments, we leverage our established comprehensive noise model to optimize the couplings through grid search and Bayesian optimization (BO) \cite{Khatri2021,Niu2019}. For $N=4$, we first conduct a grid search over $J_0 \in [0.1, 4.0]$ with a 0.1 step size, targeting the three coupling strengths $J_{1,2}, J_{2,3}, J_{3,4}$, following the formula $J_{i,i+1} = J_0 \cdot \sqrt{i (N - i)}$. The top three uniform $J_0$ values 2.9, 3.0 and 2.8. Among them, $J_0 = 2.9$ yielded peak SP in simulation, achieving 0.766 on \texttt{ibm\_sherbrooke} and 0.660 on \texttt{ibm\_brisbane}, as shown in Figure \ref{fig:Grid-Search}(a). Experimental validation on \texttt{ibm\_sherbrooke} confirmed an average peak SP improvement of 0.072($\sim$10.3\%), with high stability across runs. Even on the \texttt{ibm\_brisbane}, SP improvements are notable, validating the noise model’s ability to accurately reflect real quantum hardware performance and reduce computational costs.

To further enhance the performance, we employ BO to explore non-uniform $J_0$ configurations, relaxing the mirror symmetry constraint to prioritize SP \cite{Frazier2018}. BO utilizes a Gaussian process (GP) to model peak SP as a function of $J_{1,2}, J_{2,3}, J_{3,4}$, enforcing the constraint $J_{2,3} > J_{1,2}, J_{3,4}$ \cite{Gardner2014}. Starting from the top grid search results ($J_0 = 2.9, 3.0, 2.8$), we perform five iterations per starting point. BO incorporates adaptive sensitivity analysis by perturbing $J_0$ (increment 0.01) to estimate SP gradients, dynamically adjusting the search range
\begin{equation}
\Delta = \min\left[0.15, \max\left(0.05, \frac{0.1}{\text{sensitivity} + 10^{-6}}\right)\right].
\label{eq:Delta}
\end{equation}
This formula ensures that the search range inversely scales with parameter sensitivity: high sensitivity prompts a narrower range for precise exploration, while low sensitivity widens the range to explore potential optima. The lower bound of 0.05 prevents over-localization, avoiding suboptimal solutions, while the upper bound of 0.15 limits excessive exploration for efficiency. The $10^{-6}$ term prevents division by zero \cite{Shahriari2016}. Additionally, we introduce an attention mechanism to prioritize regions with high SP sensitivity, accelerating convergence \cite{Brochu2010}. The optimal non-uniform configuration, $J_{1,2}=2.9788, J_{2,3}=3.0182, J_{3,4}=2.8212$, achieves peak SPs of 0.781 on \texttt{ibm\_sherbrooke} and 0.696 on \texttt{ibm\_brisbane}, surpassing the uniform $J_0=2.9$ SPs of 0.766 and 0.660, as shown in Figure \ref{fig:Grid-Search}(b). This configuration maintains higher SP in the second period with reduced decay and faster transmission velocity, demonstrating that non-uniform coupling significantly enhances noise resilience.

\subsection*{Comprehensive noise model for different length of the chain}
In above analysis, we mainly consider an $N=4$ chain. How our comprehensive noise model performs for different length of the chain? We first study $N=3$ case. The noise parameters are the same as in Table~\ref{tab:Noise-Parameters}. These parameters, sourced from \texttt{ibm\_sherbrooke}'s calibration data, ensure consistency with the hardware environment.

\begin{figure}[!htb]
    \centering
    \includegraphics[width=1\columnwidth]{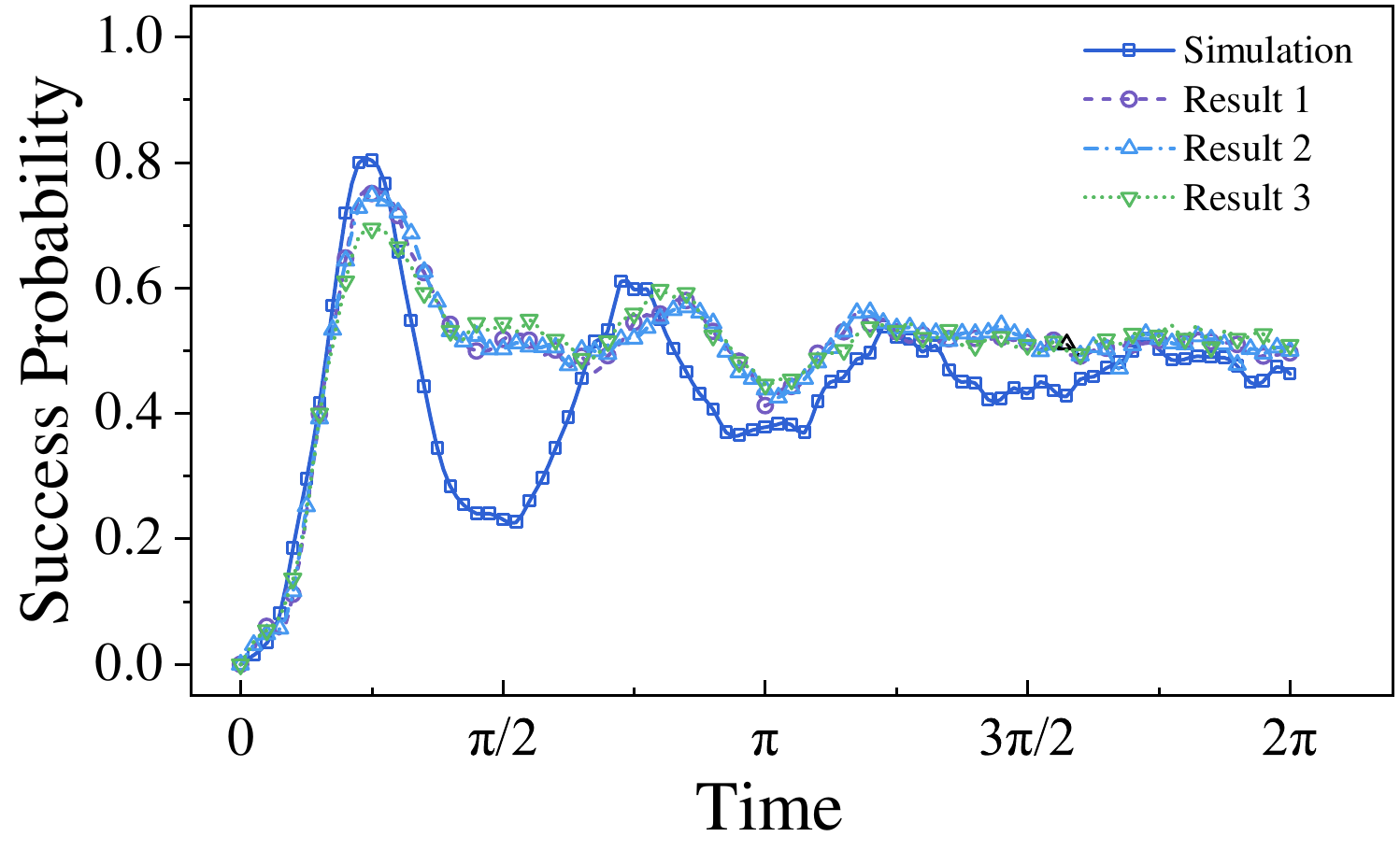}
    \caption{SP evolution under the comprehensive noise model compared to experimental runs on \texttt{ibm\_sherbrooke} for $N=3$. The solid line represents the simulated SP, while points indicate experimental data, showing a first-period peak of $\sim 0.801$ at $t^* \approx \pi/4$, a second-period peak of $\sim 0.61$, and fluctuations around 0.5 after $t = \pi$.}
    \label{fig:Comprehensive-Noise-N3}
\end{figure}

The simulation employs an 80-step Suzuki-Trotter decomposition over a total evolution time of $T = 2\pi$, with the initial state $|100\rangle$ and target state $|001\rangle$. As shown in Fig.~\ref{fig:Comprehensive-Noise-N3}(b), the simulated SP under the comprehensive noise model reaches a peak of approximately 0.801 in the first period at $t^* \approx \pi/4$, about 0.04 higher than the 0.76 peak for $N=4$. This improvement is primarily due to the shorter chain length, which reduces circuit depth and thus the cumulative effects of $T_1$/$T_2$ decoherence and gate errors. The hitting time shifts from the theoretical value $t^* = \pi/2$ to $t^* \approx \pi/4$, consistent with $N=4$, primarily driven by coherent phase evolution induced by ZZ crosstalk. Experimental results on \texttt{ibm\_sherbrooke} show an average peak SP of 0.721, higher than the average of 0.694 for \texttt{ibm\_sherbrooke} for $N=4$. Simulations demonstrate close agreement with experiments, especially around peak SP, validating the effectiveness of our comprehensive noise model.

For longer chains ($N > 4$), we believe the linear increase in circuit depth will amplify the effects of $T_1$/$T_2$ decoherence and gate errors, potentially reducing peak SP below practical thresholds, making algorithmic PST challenging. Additionally, non-Markovian noise, not fully captured in the current model, may exacerbate hitting time shifts and SP degradation. Future research could explore modeling non-Markovian effects through time-correlated noise channels and employ advanced mitigation strategies, such as dynamical decoupling \cite{de2010universal} or surface code error correction \cite{google2023suppressing}, to maintain algorithmic PST performance for longer chains.

\subsection*{Transfer of Arbitrary Quantum State}
To demonstrate the generality of our QST protocol, we analyze the transfer of an arbitrary state $|\psi\rangle = A |0\rangle +B |1\rangle $ , specifically $|\psi\rangle = \frac{1}{\sqrt{2}} \left( |0\rangle + |1\rangle \right)$ with $A=B=\frac{1}{\sqrt{2}}$, in a 4-qubit XY chain with PST couplings \cite{xiang2024enhanced}. The initial state $\frac{1}{\sqrt{2}} \left( |0\rangle + |1\rangle \right) \otimes |000\rangle$ is prepared by applying a Hadamard gate to the first qubit, followed by Suzuki-Trotter evolution over a total time $T = 2\pi$ with 40 steps. The goal is to transfer the state to the last qubit.

The success of the state transfer is quantified by reconstructing the density matrix of the last qubit through quantum state tomography~\cite{zhang2017efficient}. Measurements are performed in the $X$, $Y$, and $Z$ bases to obtain the expectation values $\langle \sigma_x \rangle$, $\langle \sigma_y \rangle$, and $\langle \sigma_z \rangle$. For the $X$ basis, a Hadamard gate is applied to the last qubit before measurement in the computational basis. For the $Y$ basis, an $S^\dagger$ gate, defined by the matrix
\begin{equation}
S^\dagger = \begin{pmatrix} 1 & 0 \\ 0 & -i \end{pmatrix},
\end{equation}
followed by a Hadamard gate, is applied before measurement~\cite{Nielsen2010}. The $Z$ basis requires no additional gates. Each measurement uses 2048 shots to estimate probabilities $p_0$ and $p_1$, from which expectation values are computed as $\langle \sigma_i \rangle = p_0 - p_1$ ($i = x, y, z$). Here, $p_0$ and $p_1$ denote the probabilities of measuring $|0\rangle$ and $|1\rangle$, respectively, in the computational basis after applying the corresponding basis transformation gates. The density matrix is reconstructed as~\cite{james2001measurement}
\begin{equation}
\rho = \frac{1}{2} \left( I + \langle \sigma_x \rangle \sigma_x + \langle \sigma_y \rangle \sigma_y + \langle \sigma_z \rangle \sigma_z \right).
\end{equation}
Using the reconstructed density matrix,  the SP can be computed by \cite{Nielsen2010}:
\begin{equation}
P = \left( tr \sqrt{\sqrt{\rho_{\text{target}}} \rho \sqrt{\rho_{\text{target}}} } \right)^2,
\end{equation}
where $\rho_{\text{target}}$ is the target density matrix.

\begin{figure}[!htb]
    \centering
    \includegraphics[width=1\columnwidth]{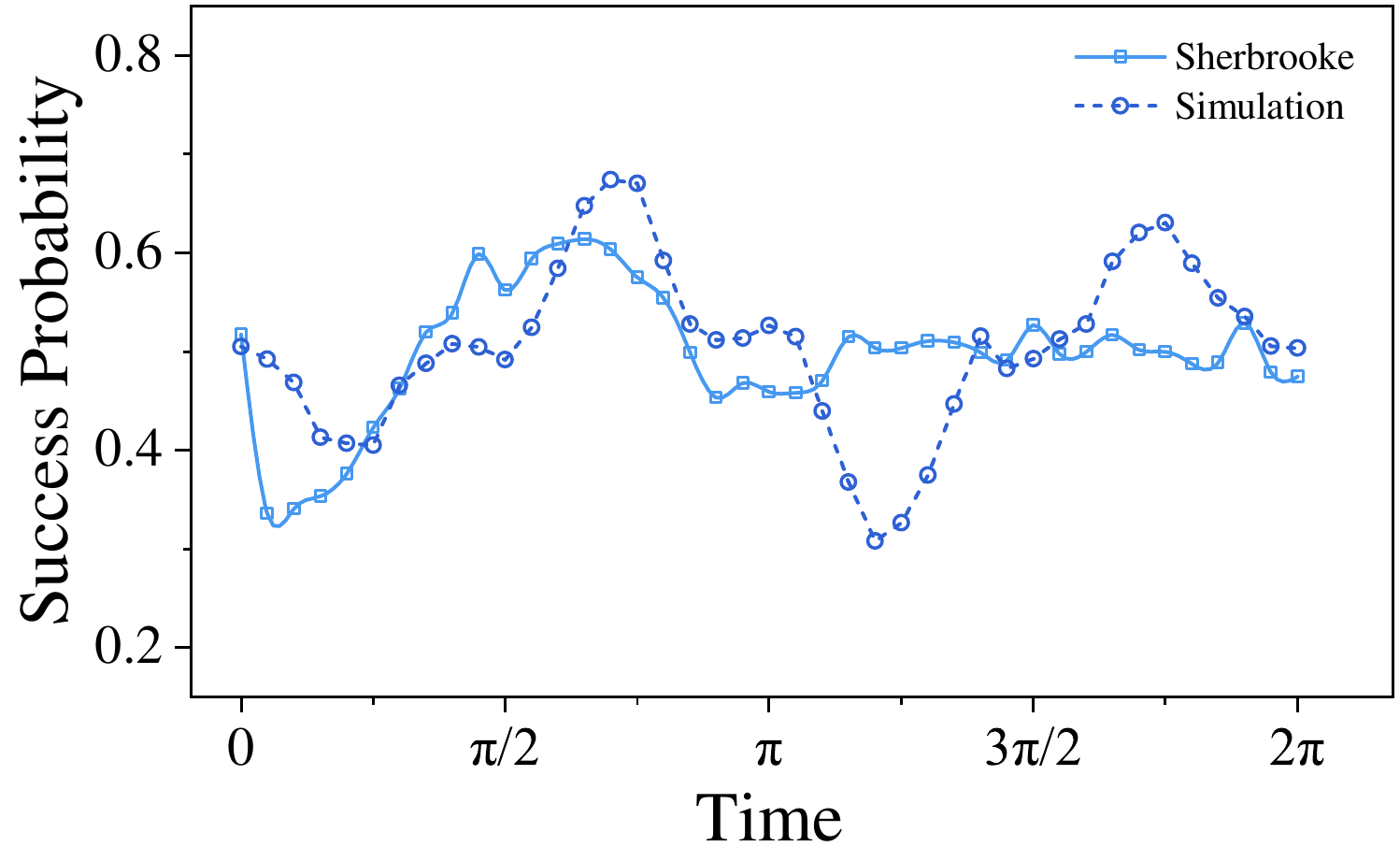}
    \caption{SP evolution for the transfer of state $\frac{1}{\sqrt{2}} \left( |0\rangle + |1\rangle \right)$, comparing simulations under the comprehensive noise model with results from \texttt{ibm\_sherbrooke} ($N=4$).}
    \label{fig:Arbitrary-Noise}
\end{figure}

Simulations using our comprehensive noise model, with parameters identical to those in Table~\ref{tab:Noise-Parameters} (including Pauli and depolarizing noise, thermal relaxation ($T_1$) and dephasing ($T_2$), ZZ crosstalk), validate its applicability to arbitrary state transfer, showing close agreement with experiments on \texttt{ibm\_sherbrooke}, as depicted in Fig.~\ref{fig:Arbitrary-Noise}. The simulated SP peaks at approximately 0.674 at $t^* \approx 1.4 \times \pi/2$, while the experimental peak reaches 0.614 at $t^* \approx 1.3 \times \pi/2$. This agreement validates the robustness of our protocol for arbitrary state transfer under realistic noise conditions. Consequently, error mitigation strategies developed for single-excitation transfer, such as rescaling and coupling optimization, are applicable to arbitrary states, and further discussion of error mitigation is omitted.

\section*{Discussions}
We have proposed an effective comprehensive noise model to successfully describe the system dynamics of spin chains simulated on IBM quantum computing devices: the \texttt{ibm\_sherbrooke} and \texttt{ibm\_brisbane} processors. By studying QST process through the chain with PST couplings, the experiments showcase that our noise model can capture the observed SP evolution, effectively reflecting the impact and relative contributions of Pauli noise, thermal relaxation ($T_1$), dephasing ($T_2$), and ZZ crosstalk. The SP evolution of the QST across multiple periods, exhibiting symmetry-consistent behavior, are systematically presented. Based on the noise model, rescaling techniques and coupling optimization are both used to significantly improve the peak SP, highlighting the robustness of the approach under realistic noisy conditions. The optimization of coupling strengths via grid search and Bayesian optimization further enhance the transmission SP, achieving a peak SP of 0.781, a $\sim$7.72\% improvement on \texttt{ibm\_sherbrooke}, demonstrating the potential of tailored coupling designs to mitigate noise \cite{Khatri2021,Niu2019}.

There are still notable limitations in current noise simulations. Experimental results show that noise leads to unstable and advanced hitting times, which may arise not only from ZZ crosstalk but also from non-Markovian noise. However, existing noise models fail to account for non-Markovian effects in quantum hardware. This type of time-correlated noise originates from memory-dependent interactions between the environment and the system, resulting in non-local temporal correlations \cite{White2020}. In PST experiments, the Suzuki-Trotter decomposition (with time step $\Delta t = T/n$) simulates XY interactions using $R_{XX}/R_{YY}$ gates. However, non-Markovian noise introduces correlations between neighboring Trotter steps, disrupting the linear decomposition of the evolution operator. These correlations cause phase errors to accumulate through environmental feedback, leading to deviations in the excitation transmission speed and shifts in the hitting times \cite{Shrikant2023}.

By quantitatively analyzing the contributions of different noise sources and optimizing coupling strengths, this work offers new insights into how noise disrupts quantum dynamics, providing valuable guidance for developing targeted noise mitigation strategies. Compared to traditional methods addressing single noise sources, the multi-noise collaborative suppression framework and non-uniform coupling optimization proposed here substantially enhance the transmission SP. Future research could focus on fine-tuning the rescaling parameters $\alpha$ and $\beta$, optimizing BO hyperparameters, extending quantum chains to $N>4$ to assess the method’s applicability to larger systems, and designing targeted QST protocols based on a deeper understanding of individual noise mechanisms and their contributions.

\section*{Methods}

\subsection*{Time Evolution and Suzuki-Trotter Decomposition}
In this paper, we employ the first-order Suzuki-Trotter expansion to discretize the system’s Hamiltonian into a sequence of quantum gates. This method decomposes the Hamiltonian into local interaction terms, enabling the construction of the time evolution operator as follows
\begin{equation}
U(\Delta t) \approx \prod_{i=0}^{N-2} e^{-i J_{i,i+1} \Delta t \sigma_i^x  \sigma_{i+1}^x } e^{-i J_{i,i+1} \Delta t \sigma_i^y  \sigma_{i+1}^y }, \label{eq:Trotter}
\end{equation}
where $\Delta t = T/n$, and we take $n = 80$ throughout. The XY interaction in the system is modeled as a sequence of two-qubit RXX and RYY gates, corresponding to the Hamiltonian in Eq.~(\ref{eq:Hamiltonian}).

\subsection*{Quantum Circuit Implementation}
For the $i$-th qubit pair ($i=1,\ldots,N-1$), the rotation angles of the $R_{XX}(\theta)$ and $R_{YY}(\theta)$ gates are $\theta = J_{i,i+1} \Delta t$.

The $R_{XX}(\theta)$ gate applies a rotation around the $XX$ axis by an angle $\theta$, and can be represented by
\begin{equation}
\setlength{\arraycolsep}{0.8pt}
R_{\!XX}(\theta) = \begin{pmatrix}
\cos\frac{\theta}{2} & 0 & 0 & -i\sin\frac{\theta}{2} \\
0 & \cos\frac{\theta}{2} & -i\sin\frac{\theta}{2} & 0 \\
0 & -i\sin\frac{\theta}{2} & \cos\frac{\theta}{2} & 0 \\
-i\sin\frac{\theta}{2} & 0 & 0 & \cos\frac{\theta}{2}
\end{pmatrix}.
\end{equation}

The corresponding $R_{YY}(\theta)$ gate applies a rotation around the $YY$ axis by an angle $\theta$,
\begin{equation}
R_{\!YY}(\theta) = \begin{pmatrix}
\cos\frac{\theta}{2} & 0 & 0 & i\sin\frac{\theta}{2} \\
0 & \cos\frac{\theta}{2} & -i\sin\frac{\theta}{2} & 0 \\
0 & -i\sin\frac{\theta}{2} & \cos\frac{\theta}{2} & 0 \\
i\sin\frac{\theta}{2} & 0 & 0 & \cos\frac{\theta}{2}
\end{pmatrix}.
\end{equation}

\begin{figure}[!htb]
    \centering
    \includegraphics[width=1\columnwidth]{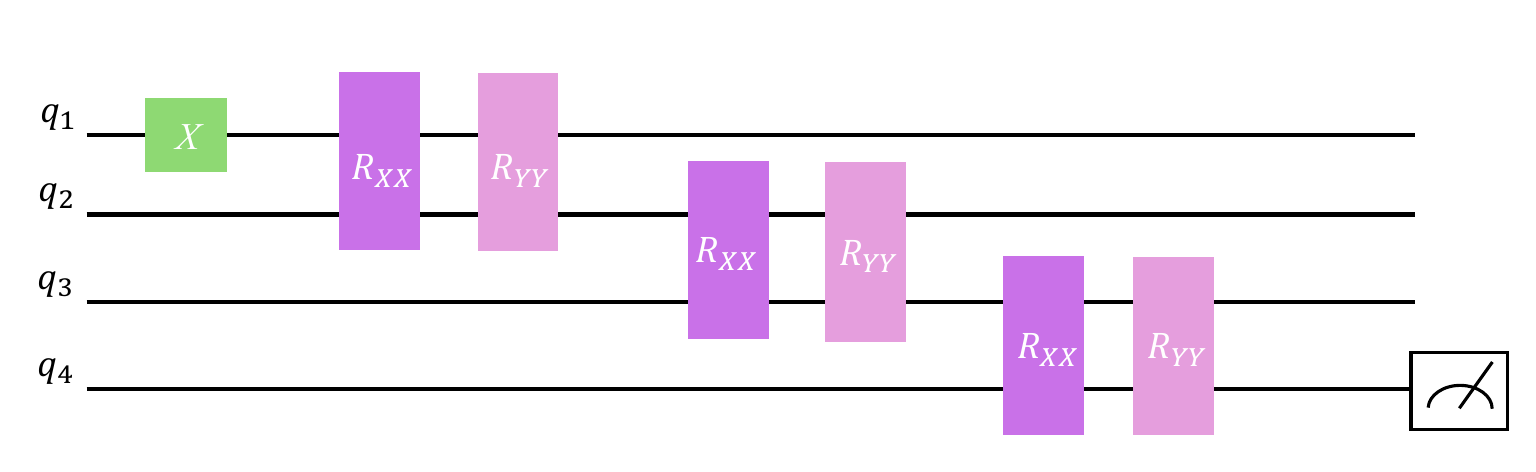}
    \caption{Circuit diagram for a single Trotter step in PST simulation ($N=4$).}
    \label{fig:Trotter-Circuit}
\end{figure}

\begin{figure}[!htb]
    \centering
    \includegraphics[width=1\columnwidth]{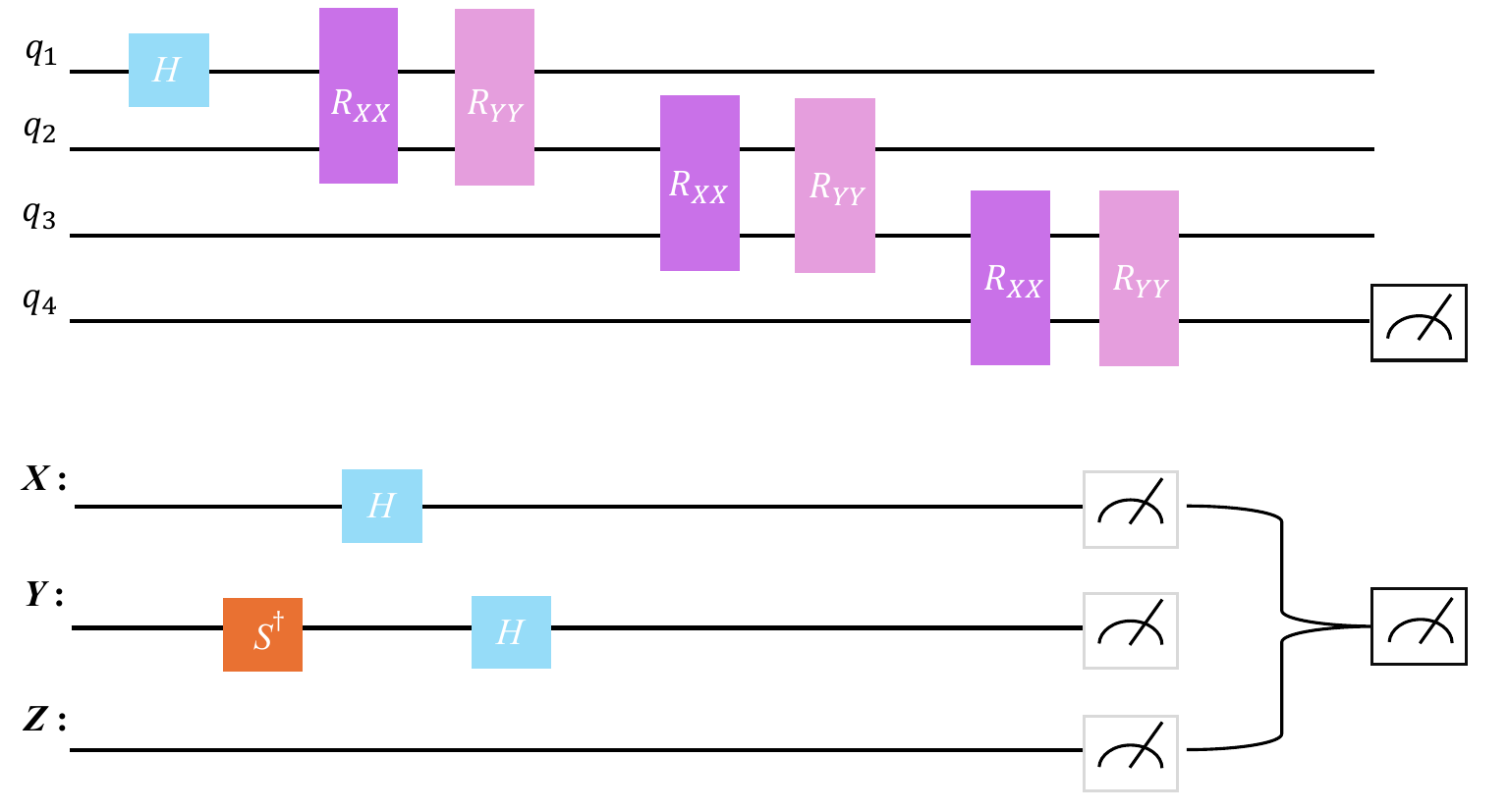}
    \caption{Circuit diagram for the arbitrary state transfer of $\frac{1}{\sqrt{2}} \left( |0\rangle + |1\rangle \right)$ in a 4-qubit chain, including measurements for quantum state tomography ($N=4$).}
    \label{fig:Arbitrary-Circuit}
\end{figure}

These gates are applied to all neighbor qubit pairs to form a single Trotter step circuit. For the single-excitation state $|1\rangle$ transfer, the total evolution is achieved by repeating this circuit 80 times. The circuit diagram for a single Trotter step is illustrated in Figure~\ref{fig:Trotter-Circuit}. For the state transfer of $|\psi\rangle = \frac{1}{\sqrt{2}} \left( |0\rangle + |1\rangle \right)$, a Hadamard gate is applied to the first qubit to prepare the initial state, followed by the Trotter evolution over 40 steps, with measurements performed in the $X$, $Y$, and $Z$ bases on the last qubit for quantum state tomography. The circuit diagram is shown in Figure~\ref{fig:Arbitrary-Circuit}.

\subsection*{Noise Model Construction}
Based on experimental data and image analysis, we construct a comprehensive noise model to understand the 20\%–30\% SP degradation in the first period for $N=4$ qubits on \texttt{ibm\_sherbrooke}. The primary noise sources include $T_1$/$T_2$ decoherence, depolarizing noise, and ZZ crosstalk. By analyzing their contributions to SP and combining circuit characteristics, we obtain reasonable values for noise parameters. $T_1$/$T_2$ decoherences, with readily available $T_1$ and $T_2$ data and gate operation times, are accurately simulated, contributing approximately 15\%–20\% to SP degradation. ZZ crosstalk, affecting phase relationships between neighboring qubits and thus the peak SP time, contributes only about 1\% to SP loss. The remaining SP degradation, approximately 15\%–20\%, is attributed to depolarizing noise, estimated by subtracting the contributions of $T_1$/$T_2$ decoherence and ZZ crosstalk from the total loss.

To validate the model, we review \texttt{ibm\_sherbrooke}’s parameters and use predicted SP degradation values to set initial noise model parameters. By studying each noise source individually, we confirm parameter reasonableness if the SP degradation range matches expectations. The comprehensive noise model’s SP degradation curve closely matches experimental results, validating its accuracy and establishing the parameters as a reasonable baseline.

\subsection*{Grid Search and Bayesian Optimization of the Coupling Strengths}
To identify optimal uniform scaling factors for the coupling strengths, we employ a grid search method \cite{Pirhooshyaran2020}. Specifically, we vary the scaling factor $J_0$ from 0.1 to 4.0 with increments of 0.1, setting the coupling strengths as $J_{i,i+1} = J_0 \cdot \sqrt{i (N - i)}$ for $i = 1, 2, 3$ and $N = 4$. For each value of $J_0$, we simulate QST under the comprehensive noise model and record the peak SP. The top three values of $J_0$ that achieve the highest SPs are 2.9, 3.0, and 2.8. These values are then used as starting points for further optimization using Bayesian Optimization.

Subsequently, to explore non-uniform coupling configurations, we utilized BO, a probabilistic approach ideal for optimizing computationally expensive functions like quantum circuit SP. BO models the peak SP as a function of individual scaling factors $J_{1,2}, J_{2,3}, J_{3,4}$ for each coupling, using a Gaussian process to predict SP for untested parameters and quantify uncertainty. Unlike grid search, BO employs an acquisition function to intelligently select parameters, balancing exploration of uncertain regions and exploitation of high-SP regions, thus reducing the number of evaluations \cite{Jones1998,Frazier2018}. We enforce the constraint $J_{2,3} > J_{1,2}, J_{3,4}$ to enhance intermediate coupling strength for noise resilience \cite{Gardner2014}. Starting from the grid search results ($J_0 = 2.9, 3.0, 2.8$), we conduct five iterations per starting point, with search ranges adjusted adaptively via sensitivity analysis (Eq. \ref{eq:Delta}). An attention mechanism prioritizes high-sensitivity regions, improving convergence efficiency greatly compared to fixed-range methods \cite{Shahriari2016,Brochu2010}. The optimal configuration, validated on \texttt{ibm\_sherbrooke}, achieves a peak SP of 0.9155, confirming the efficacy of non-uniform coupling in enhancing noise robustness.

\section*{Data and code availability}
The data generated and analyzed during this study can be made available upon reasonable request to the corresponding author. The code used to generate the numerical results presented in this paper can be made available upon reasonable request.

\section*{Acknowledgment}
We acknowledge the use of IBM quantum computers for this work. The views expressed are those of the authors and do not reflect the official policy of IBM nor the IBM quantum team. This paper is based upon work supported by the Natural Science Foundation of Shandong Province (Grant No. ZR2024MA046, ZR2021LLZ004) and Fundamental Research Funds for the Central Universities (Grant No. 202364008). L.-A. Wu has received support from the Spanish Ministry for Digital Transformation and of Civil Service of the Spanish Government through the QUANTUM ENIA project call - Quantum Spain, EU through the Recovery, Transformation and Resilience Plan-NextGenerationEU within the framework of the Digital Spain 2026.

\section*{Author contributions}
Z.-Y.G. performed the algorithm implementation, made the figures, and prepared the first version of the manuscript, Z.-M.W. initiated the project and reviewed the manuscript. All authors (Z.-Y.G., L.-A.W., and Z.-M.W.) participated in the discussions and the writing of the manuscript.

\section*{Competing interests}
The authors declare no competing interests.

\section*{Correspondence}
Correspondence and requests for materials should be addressed to Zhao-Ming Wang (email: wangzhaoming@ouc.edu.cn).

\nolinenumbers
\bibliographystyle{unsrt}
\bibliography{refs}

\begin{thebibliography}{10}

\bibitem{Nielsen2010}
Michael~A Nielsen and Isaac~L Chuang.
\newblock {\em Quantum computation and quantum information}, volume~2.
\newblock Cambridge university press Cambridge, 2001.

\bibitem{Ladd}
T.~D. Ladd, F.~Jelezko, R.~Laflamme, Y.~Nakamura, C.~Monroe, and J.~L. O'Brien.
\newblock Quantum computers.
\newblock {\em Nature}, 464(7285):45--53, 2010.

\bibitem{Shor}
Peter~W. Shor.
\newblock Algorithms for quantum computation: Discrete logarithms and
  factoring.
\newblock In {\em Proceedings of the 35th Annual Symposium on Foundations of
  Computer Science}, pages 124--134. IEEE, 1994.

\bibitem{Cao2019}
Yudong Cao, Jonathan Romero, Jonathan~P. Olson, Matthias Degroote, Peter~D.
  Johnson, M{\'a}ria Kieferov{\'a}, Ian~D. Kivlichan, Tim Menke, Borja
  Peropadre, Nicolas P.~D. Sawaya, and Al{\'a}n Aspuru-Guzik.
\newblock Quantum chemistry in the age of quantum computing.
\newblock {\em Chemical Reviews}, 119(19):10856--10915, 2019.

\bibitem{Leon}
N.~P. Leon, Z.~Bao, B.~A. Moores, L.~E. Ocola, and D.~D. Awschalom.
\newblock Materials challenges and opportunities for quantum computing
  hardware.
\newblock {\em Science}, 372(6539):eabb2823, 2021.

\bibitem{Marx}
V.~Marx.
\newblock Biology begins to tangle with quantum computing.
\newblock {\em Nature Methods}, 18(7):715--719, 2021.

\bibitem{Herman}
Dylan Herman, Cody Googin, Xiaoyuan Liu, Yue Sun, Alexey Galda, Ilya Safro,
  Marco Pistoia, and Yuri Alexeev.
\newblock Quantum computing for finance.
\newblock {\em Nature Reviews Physics}, 5(8):450--465, 2023.

\bibitem{Preskill2018}
John Preskill.
\newblock Quantum computing in the nisq era and beyond.
\newblock {\em Quantum}, 2:79, 2018.

\bibitem{Kandala2019}
Abhinav Kandala, Kristan Temme, Antonio~D. C{\'o}rcoles, Antonio Mezzacapo,
  Jerry~M. Chow, and Jay~M. Gambetta.
\newblock Error mitigation extends the computational reach of a noisy quantum
  processor.
\newblock {\em Nature}, 567(7749):491--495, 2019.

\bibitem{Nam}
Yunseong Nam, Jwo-Sy Chen, Neal~C. Pisenti, Kenneth Wright, Murphy~Yuezhen Kim,
  Brendan~J. Crowley, Dmitri Maslov, and Christopher Monroe.
\newblock Ground-state energy estimation of the water molecule on a trapped-ion
  quantum computer.
\newblock {\em npj Quantum Information}, 6(1):33, 2020.

\bibitem{schiansky2023demonstration}
Peter Schiansky, Julia Kalb, Esther Sztatecsny, Marie-Christine Roehsner,
  Tobias Guggemos, Alessandro Trenti, Mathieu Bozzio, and Philip Walther.
\newblock Demonstration of quantum-digital payments.
\newblock {\em nature communications}, 14(1):3849, 2023.

\bibitem{Georgescu2014}
I.~M. Georgescu, S.~Ashhab, and Franco Nori.
\newblock Quantum simulation.
\newblock {\em Reviews of Modern Physics}, 86(1):153--185, 2014.

\bibitem{Wuetal}
Lian-Ao Wu and Mark~S. Byrd.
\newblock Quantum simulation.
\newblock {\em Progress in Nuclear Magnetic Resonance Spectroscopy},
  59(2):81--82, 2011.

\bibitem{Smith2019}
Adam Smith, M.~S. Kim, Frank Pollmann, and Johannes Knolle.
\newblock Simulating quantum many-body dynamics on a current digital quantum
  computer.
\newblock {\em npj Quantum Information}, 5(1):106, 2019.

\bibitem{WuReport}
G.~Quiroz and D.~A. Lidar.
\newblock Dynamically generated decoherence-free subspaces and subsystems on
  superconducting qubits.
\newblock {\em Reports on Progress in Physics}, 87(9):097601, 2024.

\bibitem{Guzik}
Al{\'a}n Aspuru-Guzik, Anthony~D. Dutoi, Peter~J. Love, and Martin Head-Gordon.
\newblock Simulated quantum computation of molecular energies.
\newblock {\em Science}, 309(5741):1704--1707, 2005.

\bibitem{YunlanJi}
Yunlan Ji, Feifei Zhou, Xi~Chen, Ran Liu, Zhaokai Li, Hui Zhou, and Xinhua
  Peng.
\newblock Counterdiabatic transfer of a quantum state in a tunable heisenberg
  spin chain via the variational principle.
\newblock {\em Physical Review A}, 105(5):052422, 2022.

\bibitem{YunlanJinjp}
Yunlan Ji, Ze~Wu, Ran Liu, Yuchen Li, Fangzhou Jin, Hui Zhou, and Xinhua Peng.
\newblock Inverse engineering for robust state transport along a spin chain via
  low-energy subspaces.
\newblock {\em New Journal of Physics}, 26(1):013041, 2024.

\bibitem{Bose2003}
S.~Bose.
\newblock Quantum communication through an unmodulated spin chain.
\newblock {\em Physical Review Letters}, 91(20):207901, 2003.

\bibitem{Karbach2005}
P.~Karbach and J.~Stolze.
\newblock Spin chains as perfect quantum state mirrors.
\newblock {\em Physical Review A}, 72(3):030301, 2005.

\bibitem{Christandl2004}
M.~Christandl, N.~Datta, A.~Ekert, and A.~J. Landahl.
\newblock Perfect state transfer in quantum spin networks.
\newblock {\em Physical Review Letters}, 92(18):187902, 2004.

\bibitem{wang2013fault}
Zhao-Ming Wang, Lian-Ao Wu, Michele Modugno, Wang Yao, and Bin Shao.
\newblock Fault-tolerant almost exact state transmission.
\newblock {\em Scientific Reports}, 3(1):3128, 2013.

\bibitem{zhang2005simulation}
Jingfu Zhang, Gui~Lu Long, Wei Zhang, Zhiwei Deng, Wenzhang Liu, and Zhiheng
  Lu.
\newblock Simulation of heisenberg xy interactions and realization of a perfect
  state transfer in spin chains using liquid nuclear magnetic resonance.
\newblock {\em Physical Review A}, 72(1):012331, 2005.

\bibitem{bellec2012faithful}
M.~Bellec, G.~M. Nikolopoulos, and S.~Tzortzakis.
\newblock Faithful communication hamiltonian in photonic lattices.
\newblock {\em Optics Letters}, 37(21):4504--4506, 2012.

\bibitem{perez2013coherent}
A.~Perez-Leija, R.~Keil, A.~Kay, H.~Moya-Cessa, S.~Nolte, L.-C. Kwek, B.~M.
  Rodr{\'i}guez-Lara, A.~Szameit, and D.~N. Christodoulides.
\newblock Coherent quantum transport in photonic lattices.
\newblock {\em Physical Review A}, 87(1):012309, 2013.

\bibitem{chapman2016experimental}
R.~J. Chapman, M.~Santandrea, Z.~Huang, G.~Corrielli, A.~Crespi, M.-H. Yung,
  R.~Osellame, and A.~Peruzzo.
\newblock Experimental perfect state transfer of an entangled photonic qubit.
\newblock {\em Nature Communications}, 7:11339, 2016.

\bibitem{xiang2024enhanced}
L.~Xiang, J.~Chen, Z.~Zhu, Z.~Song, Z.~Bao, X.~Zhu, F.~Jin, K.~Wang, S.~Xu,
  Y.~Zou, et~al.
\newblock Enhanced quantum state transfer by circumventing quantum chaotic
  behavior.
\newblock {\em Nature Communications}, 15(1):4918, 2024.

\bibitem{cai2024protecting}
W.~Cai, X.~Mu, W.~Wang, J.~Zhou, Y.~Ma, X.~Pan, Z.~Hua, X.~Liu, G.~Xue, H.~Yu,
  et~al.
\newblock Protecting entanglement between logical qubits via quantum error
  correction.
\newblock {\em Nature Physics}, 20(6):1022--1028, 2024.

\bibitem{zhang2024mech}
H.~Zhang, K.~Yin, A.~Wu, H.~Shapourian, A.~Shabani, and Y.~Ding.
\newblock Mech: Multi-entry communication highway for superconducting quantum
  chiplets.
\newblock {\em Proceedings of the 29th ACM International Conference on
  Architectural Support for Programming Languages and Operating Systems},
  2:699--714, 2024.

\bibitem{IBMQ}
{IBM Quantum}.
\newblock Ibm quantum system resources.
\newblock \url{https://quantum.cloud.ibm.com/computers}, 2025.
\newblock Accessed: July 7, 2025.

\bibitem{Salimi2013}
S.~Salimi, S.~Ghoraishipour, and A.~Sorouri.
\newblock Perfect state transfer via quantum probability theory.
\newblock {\em Quantum Information Processing}, 12(1):505--523, 2013.

\bibitem{christandl2005perfect}
Matthias Christandl, Nilanjana Datta, Tony~C Dorlas, Artur Ekert, Alastair Kay,
  and Andrew~J Landahl.
\newblock Perfect transfer of arbitrary states in quantum spin networks.
\newblock {\em Physical Review A—Atomic, Molecular, and Optical Physics},
  71(3):032312, 2005.

\bibitem{Yung2005}
Man-Hong Yung and Sougato Bose.
\newblock Perfect state transfer, effective gates, and entanglement generation
  in engineered bosonic and fermionic networks.
\newblock {\em Physical Review A}, 71(3):032310, 2005.

\bibitem{QiskitAer}
{Qiskit Aer Developers}.
\newblock Qiskit aer.
\newblock \url{https://github.com/Qiskit/qiskit-aer}, 2025.
\newblock Accessed: May 15, 2025.

\bibitem{Geller2013}
Michael~R. Geller and Zhongyuan Zhou.
\newblock Efficient error models for fault-tolerant architectures and the pauli
  twirling approximation.
\newblock {\em Physical Review A}, 88(1):012314, 2013.

\bibitem{Cheng2021}
Sichun Cheng, Rui Chen, and Yong Yang.
\newblock Simulating noisy quantum circuits with matrix product density
  operators.
\newblock {\em Physical Review Research}, 3(2):023005, 2021.

\bibitem{Temme2017}
Kristan Temme, Sergey Bravyi, and Jay~M. Gambetta.
\newblock Error mitigation for short-depth quantum circuits.
\newblock {\em Physical Review Letters}, 119(18):180509, 2017.

\bibitem{QiskitRuntime}
{IBM Quantum}.
\newblock Qiskit runtime documentation.
\newblock \url{https://docs.quantum.ibm.com/run}, 2025.
\newblock Accessed: May 15, 2025.

\bibitem{Li2017}
Ying Li and Simon~C. Benjamin.
\newblock Efficient variational quantum simulator incorporating active error
  minimization.
\newblock {\em Physical Review X}, 7(2):021050, 2017.

\bibitem{Kandala2017}
Abhinav Kandala, Antonio Mezzacapo, Kristan Temme, Maika Takita, Markus Brink,
  Jerry~M. Chow, and Jay~M. Gambetta.
\newblock Hardware-efficient variational quantum eigensolver for small
  molecules and quantum magnets.
\newblock {\em Nature}, 549(7671):242--246, 2017.

\bibitem{Babukhin2022}
D.~V. Babukhin and W.~V. Pogosov.
\newblock The effect of quantum noise on algorithmic perfect quantum state
  transfer on nisq processors.
\newblock {\em Quantum Information Processing}, 21(1):7, 2022.

\bibitem{Flammia2020}
Steven~T. Flammia and Joel~J. Wallman.
\newblock Efficient estimation of pauli channels.
\newblock {\em ACM Transactions on Quantum Computing}, 1(1):1--32, 2020.

\bibitem{Chen2023}
Senrui Chen, Yunchao Liu, Matthew Otten, and Alireza Seif.
\newblock The learnability of pauli noise.
\newblock {\em Nature Communications}, 14(1):52, 2023.

\bibitem{Urbanek2021}
Miroslav Urbanek, Benjamin Nachman, and Wibe~A. de~Jong.
\newblock Mitigating depolarizing noise on quantum computers with
  noise-estimation circuits.
\newblock {\em Physical Review Letters}, 127(27):270502, 2021.

\bibitem{Escofet2025}
P.~Escofet, S.~Filipp, and M.~M{\"u}ller.
\newblock An accurate and efficient analytic model of fidelity under
  depolarizing noise oriented to large scale quantum system design, 2025.

\bibitem{Gonzalez-Garcia2025}
G.~Gonz{\'a}lez-Garc{\'i}a, J.~I. Cirac, and R.~Trivedi.
\newblock Pauli path simulations of noisy quantum circuits beyond average case.
\newblock {\em Quantum}, 9:1730, 2025.

\bibitem{Brown2024}
A.~F. Brown and D.~A. Lidar.
\newblock Efficient chromatic-number-based multi-qubit decoherence and
  crosstalk suppression, 2024.

\bibitem{McKay2023}
David~C. McKay, Akel T.~K. Hashim, Seth Merkel, Carlos Ferrando-Soria, and
  Irfan Siddiqi.
\newblock Benchmarking quantum processor performance at scale, 2023.
\newblock arXiv preprint.

\bibitem{Magesan2020}
Easwar Magesan and Jay~M. Gambetta.
\newblock Effective hamiltonian models of the cross-resonance gate.
\newblock {\em Physical Review A}, 101(5):052308, 2020.

\bibitem{Mitchell2021}
B.~K. Mitchell, R.~K. Naik, Akel Hashim, J.~M. Kreikebaum, David Santiago,
  Irfan Siddiqi, and Andrew T.~K. Jordan.
\newblock Hardware-efficient microwave-activated tunable coupling between
  superconducting qubits.
\newblock {\em Physical Review Letters}, 127(20):200502, 2021.

\bibitem{Ni2022}
Zhantao Ni, Sai Li, Xing Deng, Yuxuan Wu, Libo Hu, and Jian Li.
\newblock Scalable method for eliminating residual zz interaction between
  superconducting qubits.
\newblock {\em Physical Review Letters}, 129(4):040502, 2022.

\bibitem{Krinner2022}
S.~Krinner, N.~Lacroix, C.~Hellings, A.~Remm, C.~K. Andersen, G.~Salis,
  M.~Kjaergaard, and A.~Wallraff.
\newblock Realizing repeated quantum error correction in a distance-three
  surface code.
\newblock {\em Nature}, 605(7911):669--674, 2022.

\bibitem{Tripathi2022}
Vinay Tripathi, Hao-Kai Xu, C.~R. Conner, R.~McDermott, and M.~Saffman.
\newblock Operation of a silicon quantum processor unit cell above one kelvin.
\newblock {\em Nature}, 601(7893):350--354, 2022.

\bibitem{Strikis2021}
Armands Strikis, Dayou Qin, Yanzhu Chen, Simon~C. Benjamin, and Ying Li.
\newblock Learning-based quantum error mitigation.
\newblock {\em PRX Quantum}, 2(4):040330, 2021.

\bibitem{Khatri2021}
Sumeet Khatri, Ryan LaRose, Alexander T.~K. Poremba, Lukasz Cincio, Andrew~T.
  Sornborger, and Patrick~J. Coles.
\newblock Quantum-assisted quantum compiling.
\newblock {\em PRX Quantum}, 2(1):010324, 2021.

\bibitem{Niu2019}
Murphy~Yuezhen Niu, Sergio Boixo, Vadim~N. Smelyanskiy, and Hartmut Neven.
\newblock Universal quantum control through deep reinforcement learning.
\newblock {\em npj Quantum Information}, 5(1):33, 2019.

\bibitem{Frazier2018}
Peter~I. Frazier.
\newblock A tutorial on bayesian optimization, 2018.

\bibitem{Gardner2014}
Jacob~R. Gardner, Matt~J. Kusner, Zhixiang~E. Xu, Kilian~Q. Weinberger, and
  John~P. Cunningham.
\newblock Bayesian optimization with inequality constraints.
\newblock {\em Proceedings of the 31st International Conference on Machine
  Learning}, 32:937--945, 2014.

\bibitem{Shahriari2016}
Bobak Shahriari, Kevin Swersky, Ziyu Wang, Ryan~P. Adams, and Nando de~Freitas.
\newblock Taking the human out of the loop: A review of bayesian optimization.
\newblock {\em Proceedings of the IEEE}, 104(1):148--175, 2016.

\bibitem{Brochu2010}
Eric Brochu, Vlad~M. Cora, and Nando de~Freitas.
\newblock A tutorial on bayesian optimization of expensive cost functions, with
  application to active user modeling and hierarchical reinforcement learning,
  2010.

\bibitem{de2010universal}
Gijs De~Lange, Zhi-Hui Wang, D~Riste, VV~Dobrovitski, and R~Hanson.
\newblock Universal dynamical decoupling of a single solid-state spin from a
  spin bath.
\newblock {\em Science}, 330(6000):60--63, 2010.

\bibitem{google2023suppressing}
google2023suppressing.
\newblock Suppressing quantum errors by scaling a surface code logical qubit.

\bibitem{zhang2017efficient}
Jiaojiao Zhang, Kezhi Li, Shuang Cong, and Haitao Wang.
\newblock Efficient reconstruction of density matrices for high dimensional
  quantum state tomography.
\newblock {\em Signal Processing}, 139:136--142, 2017.

\bibitem{james2001measurement}
Daniel~FV James, Paul~G Kwiat, William~J Munro, and Andrew~G White.
\newblock Measurement of qubits.
\newblock {\em Physical Review A}, 64(5):052312, 2001.

\bibitem{White2020}
G.~A.~L. White, C.~D. Hill, J.~Pollock, L.~C.~L. Hollenberg, and K.~Modi.
\newblock Demonstration of non-markovian process characterisation and control
  on a quantum processor.
\newblock {\em Nature Communications}, 11(1):6301, 2020.

\bibitem{Shrikant2023}
U.~Shrikant and Prabha Mandayam.
\newblock Quantum non-markovianity: Overview and recent developments.
\newblock {\em Frontiers in Quantum Science and Technology}, 2:1134583, 2023.

\bibitem{Pirhooshyaran2020}
Mohammad Pirhooshyaran and Tam{\'a}s Terlaky.
\newblock Quantum circuit design search, 2020.

\bibitem{Jones1998}
Donald~R. Jones, Matthias Schonlau, and William~J. Welch.
\newblock Efficient global optimization of expensive black-box functions.
\newblock {\em Journal of Global Optimization}, 13(4):455--492, 1998.

\end{thebibliography}

\end{document}